\documentclass[twocolumn]{article}
\usepackage[utf8]{inputenc}
\usepackage{amsmath}
\usepackage{mathtools, cuted}
\usepackage{caption}
\usepackage[dvipsnames]{xcolor}
\usepackage{graphicx}
\usepackage{braket}
\usepackage{booktabs}
\usepackage{soul}
\usepackage{cancel}
\usepackage{abstract}
\usepackage[normalem]{ulem}

\usepackage[colorlinks=true, urlcolor=MidnightBlue, linkcolor=MidnightBlue, 
citecolor=Mahogany, pdfborder={0 0 0}]{hyperref}
\usepackage[sort&compress,numbers]{natbib}
\usepackage{doi}
\usepackage[capitalise]{cleveref}

\title{\Large\bfseries{The phase of the electromagnetic form factor of the pion}}
\author{\normalsize{
        Enrique Ruiz Arriola{\color{Mahogany}\thanks{earriola@ugr.es}}$^{\color{Mahogany}{\ a,b}}$, 
        Pablo Sanchez-Puertas{\color{Mahogany}\thanks{pablosanchez@ugr.es}}$^{\color{Mahogany}{\ a}}$
        }\vspace{0.2cm}\\
        {\small{$^{\color{Mahogany}{a}}$\textit{Departamento de F{\'i}sica At{\'o}mica, Molecular y Nuclear}}}\\
        {\small{$^{\color{Mahogany}{b}}$\textit{Instituto Carlos I de F{\'i}sica Te{\'o}rica y Computacional}}}\\
        {\small{\textit{Universidad de Granada, E-18071, Granada, Spain}}}
}
\date{}

\begin{document}
\renewcommand{\abstractname}{\vspace{-\baselineskip}}


\twocolumn[
  \begin{@twocolumnfalse}
    \maketitle \hrule
    \begin{abstract}
      We employ a dispersion relation that allows to recover the phase of the electromagnetic 
      form factor of the pion from its absolute value above threshold. Compared to alternative 
      approaches building on the phase, this approach builds on experimental input directly 
      accessible at colliders. Employing the precise datasets from the $e^+e^-\to\pi^+\pi^-$ 
      reaction, we obtain the phase of the electromagnetic form factor up to $2.5$~GeV, well 
      beyond standard dispersive approaches. In addition, we separate the isovector and isoscalar 
      components, that allows to extract the $P$-wave $\pi\pi$ phase shift. We also provide 
      relevant results, including the radius of the form factor and bounds in the spacelike 
      region. Last, but not least, the study assess potential systematic uncertainties from 
      the interpolation method and potential zeros of the form factor.  \\\hrule\vspace{5mm}
    \end{abstract}
  \end{@twocolumnfalse}
]
\saythanks

\section{Introduction}

The electromagnetic form factor of the pion encapsulates important
aspects of hadron dynamics. In particular, it enters in the dispersive
reconstruction of a variety of hadronic electromagnetic form factors,
see for instance
Refs.~\cite{Blatnik:1978wj,Hanhart:2013vba,Moussallam:2013una,Hoferichter:2014vra,
Colangelo:2015ama,Hoferichter:2016duk,Alarcon:2018irp,Danilkin:2018qfn,Stamen:2022uqh,Holz:2022hwz}. 
Furthermore, by virtue of Watson's final-state theorem, its phase identifies below
inelasticities with the $P$-wave $\pi\pi$ phase shift modulo
isospin-breaking (IB) corrections. Such a phase is 
ubiquitous in many hadronic quantities at low energies once the
machinery of dispersion relations is employed. In summary, this form
factor encodes relevant universal effects in low-energy QCD, and
provides as such valuable information for hadronic physics.

While the absolute value of the pion form factor can be accessed at
$e^+e^-$ colliders in the $e^+e^- \to \pi^+\pi^-$ reaction in the
time-like region, its phase is not directly measurable. Nonetheless such a
phase can be inferred using long-known dispersion relations involving
the modulus of the form factor along the unitarity 
cut~\cite{Truong:1968,Truong:1969gr,Geshkenbein:1969bb,Cronstrom:1973wr,Geshkenbein:1988ef,Geshkenbein:1998gu}
(see also \cite{Bowcock:1968txb,Okubo:1973idq,Leutwyler:2002hm}).
Seemingly, this possibility to obtain the $P$-wave $\pi\pi$ phase
shift has fallen into oblivion with the advent of the modern and
precise extractions based on Roy
equations~\cite{Ananthanarayan:2000ht,Garcia-Martin:2011iqs,Caprini:2011ky}. Motivated by
the dense and precise data that is available from modern $e^+e^- \to
\pi^+\pi^-$ analysis, we reconsider such possibility with the aid of
modulus dispersion relations that have an improved convergence with
respect to previous studies, carefully assessing on convergence 
and the impact of potential zeroes of the form factor.
Note that this study provides a unique opportunity to access the region
above inelastic thresholds up to $\sqrt{s}=3$~GeV, where the phase of
the form factor no longer identifies with that of the $P$-wave
$\pi\pi$ scattering one, and where information is scarce. Compared to
traditional dispersive approaches, that become intractable beyond the
elastic region, our approach finds itself in a privileged position to
analyze the high-energy region. In this work, we make use of the
widely-adopted Gounaris-Sakurai parameterization as an auxiliary means
to interpolate and fit the available data, including a study on 
systematics from the interpolation method. As a result, we find good agreement
when comparing to the extraction based on Roy equations, suggesting
future studies along these lines. For completeness, we also study
further applications, including the extrapolation to the spacelike
region, the pion charge radius and the  separation of  
the $P$-wave $\pi\pi$ phase shift from the octet form factor, 
clarifying some misunderstandings regarding isospin-breaking effects.

The paper is structured as follows: \cref{sec:defs} 
introduces the main definitions and dispersion relations formulae.  
Their convergence properties are analyzed by means of a toy model in 
\cref{sec:toymodel} to discern their applicability. Real data is 
analyzed in \cref{sec:realdata}, that embodies the main results in 
this work. These includes the phase of the electromagnetic form factor, 
the $P$-wave $\pi\pi$ phase shift, the isovector form factor, the 
charge radius, and the spacelike behavior. 
We draw our conclusions in \cref{sec:conclusions} and provide our 
numerical results for the form factor in \cref{app:datasets}. 
Extensive discussions on systematics are found 
in \cref{app:zeros,app:syst,app:argumentTH}.

\section{Definitions \label{sec:defs}}

The charged pion electromagnetic form factor is 
defined in terms of the matrix element of the pions with the electromagnetic 
current $J_Q^{\mu} = \sum_q Q_q\bar{q}\gamma^{\mu}q$
\begin{equation}
  \bra{\pi^+(p')} J_Q^{\mu}(0) \ket{\pi^+(p)}  = F_Q^{\pi}(q^2) (p+p')^{\mu},
\end{equation}
where $q=p'-p$ is the momentum transfer and $Q_q$ stands for the
charge of the quark $q$. For definiteness we will take as usual the notation $q^2 = -Q^2 < 0$
for space-like momenta and $q^2 = s > 0$ for time-like momenta, which
corresponds to the $e^+ e^-$ invariant center-of-mass energy squared. While such a
non-perturbative function is in general unknown, some properties are
well established. At low energies, the Ward identities imply
$F_Q^{\pi}(0)=1$ while, on the opposite extreme, at asymptotically
large Euclidean momenta $Q^2 \to \infty$, perturbative QCD
(pQCD) demands that (see
\cite{Radyushkin:1977gp,Farrar:1979aw,Efremov:1978rn,Lepage:1979zb,Efremov:1979qk,Lepage:1980fj,Melic:1998qr,Chen:2023byr})
\begin{multline}
    \lim_{Q^2\to\infty}F_{Q}^{\pi}(q^2=-Q^2) = \frac{16\pi F_{\pi}^2\alpha_s(\mu_R^2)}{Q^2} \\ 
    \times \left( 1  +\frac{\alpha_s(\mu_R^2)}{\pi} \left[ 6.58+\frac{9}{4}\ln\left(\frac{\mu_R^2}{Q^2}\right) \right] \right)
\end{multline}
for the asymptotic distribution amplitude, with $F_{\pi}\simeq 92$~MeV
the pion decay constant. Nonetheless, the scale where pQCD applies is
unclear and likely at energies not yet accessible, see for
instance~\cite{Braun:1994ij,Melic:1998qr,RuizArriola:2008sq,Bakulev:2009ib,Ananthanarayan:2018nyx,Cheng:2020vwr}
and references therein. In this limit, the analytic continuation in 
the complex $q^2$-plane from the space-like $(Q^2>0)$ 
to the time-like $(s>0)$ region corresponds to
diminish the phase by $\pi$ so that $ \ln (Q^2/\Lambda^2) \to \ln (s
e^{-i \pi}/\Lambda^2) = \ln (s/\Lambda^2) - i \pi$. Note in addition that
duality implies in the timelike region (assuming for the moment
$\mu_R^2 = Q^2$)
\begin{multline}
    \lim_{s\to\infty}F_{Q}^{\pi}(s) \to -\frac{16\pi F_{\pi}^2}{s} \frac{4\pi}{\beta_0} \frac{L +i\pi}{L^2+\pi^2} \\ 
    \left[1 +   \frac{6.58}{\pi}\frac{4\pi}{\beta_0}\frac{L +i\pi}{L^2+\pi^2} \right], \qquad 
 \end{multline} 
with $L = \ln\left(\frac{s}{\Lambda^2}\right)$, and
where we used the LO result $\alpha_s(Q^2)=(4\pi/\beta_0)/\ln(Q^2/\Lambda^2)$ 
with $\beta_0 = (11/3)N_c -(2/3)N_f$ and $\Lambda\simeq 250$~MeV.
This implies that real and imaginary parts are
negative modulo duality violations. Further, $\tan\delta = \pi/L$,
that requires the phase to behave asymptotically, modulo $2\pi$, 
as $\delta \to\pi(1+L^{-1} +[6.58/\pi][4\pi/\beta_0]L^{-2})$. 
Note this implies, at
$s=(2.5\textrm{GeV})^2, F_{Q}^{\pi} = -0.017 -0.018i$, with
corresponding phase $226^{\circ}$, see also Refs.~\cite{Gousset:1994yh,Chen:2018tch} 
for further discussions on the pQCD analytic continuation). 
As we shall discuss, while the data
shows a clear departure from pQCD (similar to the spacelike data), it
agrees qualitatively with pQCD expectations. 
One may wonder at this point whether pQCD could predict the phase 
without $2\pi$ ambiguities. In this regard, the argument theorem~\cite{Schaum}
comes in handy. Essentially, applying Cauchy's integral theorem to the 
logarithmic derivative $d \ln F_{Q}^{\pi}(s)/ds$ and choosing an appropriate 
contour to exclude singularities, one obtains (see also \cref{app:argumentTH})\footnote{In general, if pQCD 
predicts $f(s) \propto s^{-n}\alpha_s^m$, the principle of the argument 
demands $\delta(s) = \pi( n +m L^{-1} )$.} 
\begin{equation}\label{eq:argumentP}
  \delta(s) = \pi \left[ 1 +L^{-1} +\frac{6.58}{\pi}\frac{4\pi}{\beta_0}L^{-2} +N -P \right]
\end{equation} 
where $s$ should be large enough to apply pQCD and where $N(P)$ corresponds to the number 
of zeros(poles). This resembles Levinson's theorem for scattering, while 
in this case the absence of bound states imply $P=0$. As a consequence, 
additions of $2\pi$ to the pQCD prediction requires the presence of 
complex-conjugate zeros. 

On the other hand, analyticity and unitarity require the
Schwarz reflection principle to hold, implying that above the
lowest-lying threshold\footnote{We take $s_{th}=4m_{\pi}^2$. In
  principle, the presence of QED lowers the lowest threshold to the
  $\pi^0\gamma$ state, but this is negligible compared to the dominant
  $\pi^+\pi^-$ state.} $F_Q^{\pi}(s\pm i\epsilon) =
|F_Q^{\pi}(s)|e^{\pm i\delta(s)}$.  In the following, we shall assume
that the form factor has no zeroes on the first Riemann sheet (cf. the
study in \cite{Ananthanarayan:2011xt} and references therein, as well as Ref.~\cite{Leutwyler:2002hm},
\cref{app:zeros} and our comments in \cref{sec:testSR} below). Under such an
assumption, one can write Cauchy's theorem for $\ln
F_Q^{\pi}(s)$. Including a subtraction at $s=0$ to ensure convergence,
one arrives at the widely used Omn{\`e}s-like solution
\begin{equation}\label{eq:OmnesSubt}
    F_Q^{\pi}(s) = \operatorname{exp}\left( 
    \frac{s}{\pi} \int_{s_{th}}^{\infty} dz \frac{\delta(z) }{z(z-s)}
    \right)
\end{equation}
which phase, $\delta(s)$, is usually identified in the elastic region 
with the $\pi\pi$ $P$-wave phase shift $\delta_1^1(s)$ due to Watson's 
theorem~\cite{Watson:1954uc} (this holds modulo IB 
corrections, that we discuss in \cref{sec:phaseshiftOctet}). 
A much less exploited relation arises from applying Cauchy's theorem to 
$\ln F_Q^{\pi}(s)/\sqrt{s_{th}-s}$, see 
Refs.~\cite{Truong:1968,Truong:1969gr,Geshkenbein:1969bb,Cronstrom:1973wr,Geshkenbein:1988ef,Geshkenbein:1998gu},
\begin{equation}\label{eq:DR0}
    F_Q^{\pi}(s) = 
    \operatorname{exp} \left(
    \frac{\sqrt{s_{th} -s}}{\pi}  \int_{s_{th}}^{\infty}dz 
    \frac{\ln|F_Q^{\pi}(z)|}{\sqrt{z -s_{th}}(z-s)}
    \right).
\end{equation}
The expression above allows to extrapolate the form factor to $s<s_{th}$ 
from its knowledge along the cut, but also to obtain its phase,
\begin{multline}
    \delta(s) = 
    -\frac{\sqrt{s -s_{th}}}{\pi}  \operatorname{PV}\int_{s_{th}}^{\infty} dz \frac{\ln|F_Q^{\pi}(z)|}{(z-s)\sqrt{z -s_{th}}}
    \\ = 
    -\frac{\sqrt{s -s_{th}}}{\pi}  \operatorname{PV}\int_{s_{th}}^{\infty} dz \frac{\ln|F_Q^{\pi}(z)/F_Q^{\pi}(s_{th})|}{(z-s)\sqrt{z -s_{th}}},
\end{multline}
where the last line follows from \cite{Cronstrom:1973wr}. Note that the 
dispersive integral in the last line is finite at threshold. Indeed, 
the value of the form factor at zero $F_Q^{\pi}(0)=1$, together with the 
$P$-wave demanding $\delta(s) \sim (s -s_{th})^{3/2}$, imply the following 
sum rules~\cite{Cronstrom:1973wr}
\begin{align}\label{eq:sumrules}
    \operatorname{exp}\left( \frac{\sqrt{s_{th}}}{\pi} \int_{s_{th}}^{\infty}dz 
    \frac{\ln|F_Q^{\pi}(z)|}{z\sqrt{z -s_{th}}} \right) = 1, \\
    \frac{2m_{\pi}}{\pi}\int_{s_{th}}^{\infty}dz 
    \frac{\ln|F_Q^{\pi}(z)/F_Q^{\pi}(s_{th})|}{(z -s_{th})^{3/2}} = 0.
\end{align}
The first can also be inferred from the asymptotic behavior, and also
the second if considering the dispersion relation in \cref{eq:SubtTh} below,
see for instance \cite{Leutwyler:2002hm}. Unfortunately, due to the
slow convergence rate of the sum rules above, their applicability
turns out to be purely academic (see discussion in the following
section). As such, these should not be applied to draw any conclusion
about possible zeroes of the form factor as done in
Ref.~\cite{Cronstrom:1973wr}.  Indeed, to have better convergence, it
is beneficial to use further subtracted dispersion
relations. Specifically, we will use the one subtracted at zero
\cite{Geshkenbein:1998gu}
\begin{align}
    F_{Q}^{\pi}(s) &{}= \operatorname{exp}\Big(
    \frac{s\sqrt{s_{th} -s}}{\pi}\int_{s_{th}}^{\infty}dz 
    \frac{\ln|F_Q^{\pi}(z)|}{z\sqrt{z -s_{th}}(z-s)}
    \Big), \label{eq:SubtZero} \\
    \delta(s) &{}=
    -\frac{s\sqrt{s -s_{th}}}{\pi} \operatorname{PV}\int_{s_{th}}^{\infty}dz 
    \frac{\ln|F_Q^{\pi}(z)|}{z\sqrt{z -s_{th}}(z-s)},\label{eq:SubtZeroPhase}
\end{align}
that we will refer to as DR1 in the following, and the one that is 
obtained by applying Cauchy's theorem to 
$\ln [F_Q^{\pi}(s)/F_Q^{\pi}(s_{th})] /(s -s_{th})^{3/2}$~\cite{Leutwyler:2002hm},\footnote{This 
is enabled thanks to the $P$-wave nature, that demands 
$F_Q^{\pi}(s) = F_Q^{\pi}(s_{th})(1 + \alpha_1(s-s_{th})+ i\beta_{3/2}(s-s_{th})^{3/2} +... )$, 
while the presence of a $\sqrt{s -s_{th}}$ term would invalidate it.}
\begin{align}
    &F_Q^{\pi}(s)  = 
    F_Q^{\pi}(s_{th})^{1 - \left(\frac{s_{th} -s}{s_{th}}\right)^{3/2} }\times\label{eq:SubtThZero}\\ & \operatorname{exp} \Bigg[
    -\frac{s(s_{th} -s)^{3/2}}{\pi}  \int_{s_{th}}^{\infty}dz 
    \frac{\ln|F_Q^{\pi}(z)/F_Q^{\pi}(s_{th})|}{z(z -s_{th})^{3/2}(z-s)}
    \Bigg], \nonumber \\
    &\delta(s) = 
    -\ln  F_Q^{\pi}(s_{th}) \left(\frac{s -s_{th}}{s_{th}}\right)^{3/2} \label{eq:SubtThZeroPhase}\\&
    -\frac{s(s -s_{th})^{3/2}}{\pi}  \operatorname{PV}\int_{s_{th}}^{\infty}dz 
    \frac{\ln|F_Q^{\pi}(z)/F_Q^{\pi}(s_{th})|}{z(z -s_{th})^{3/2}(z-s)}, \nonumber
\end{align}
that we will refer to as DR2 in the following, and has not been used in 
previous studies despite its superior convergence properties. Note also 
that in its unsubtracted version 
\begin{multline}
    F_Q^{\pi}(s) = 
    F_Q^{\pi}(s_{th})\operatorname{exp} \Big( \\
    \frac{-(s_{th} -s)^{3/2}}{\pi}  \int_{s_{th}}^{\infty}dz 
    \frac{\ln|F_Q^{\pi}(z)/F_Q^{\pi}(s_{th})|}{(z -s_{th})^{3/2}(z-s)}
    \Big),\label{eq:SubtTh}
\end{multline}
the normalization at zero demands the following sum rule (see for instance \cite{Leutwyler:2002hm})
\begin{equation}
    \frac{s_{th}^{3/2}}{\pi \ln F_Q^{\pi}(s_{th})} 
    \int_{s_{th}}^{\infty}dz 
    \frac{\ln|F_Q^{\pi}(z)/F_Q^{\pi}(s_{th})|}{z(z -s_{th})^{3/2}} =1. \label{eq:TheSumRule}
\end{equation}
This expression embodies a fast convergence behavior at high energies
and will be relevant in our analysis to be explained
shortly. \cref{eq:SubtZero,eq:SubtZeroPhase,eq:SubtThZero,eq:SubtThZeroPhase}
and \cref{eq:TheSumRule} are the relevant equations to be used in the
following under the assumption that the form factor has no zeros.
Finally, we emphasize that the sum rule in \cref{eq:TheSumRule} plays
an important role. For instance, when extrapolating the form factor into the deep
spacelike region for asymptotically large $Q^2$,
\cref{eq:SubtTh} implies
\begin{equation}\label{eq:SRinf2}
    \frac{-(Q^2 +s_{th})^{3/2}}{\ln\left(\frac{F_{Q}^{\pi}(-Q^2)}{F_{Q}^{\pi}(s_{th})}\right)} \frac{1}{Q^2} \int_{s_{th}}^{\infty} dz \frac{\ln\Big|\frac{F_Q^{\pi}(z)}{F_Q^{\pi}(s_{th})} \Big|}{(z -s_{th})^{3/2}} \to 1 .
\end{equation}
This, together with \cref{eq:TheSumRule}, allows to express the asymptotic 
euclidean limit of DR2 as
\begin{equation}
    F_{Q}^{\pi}(-Q^2) \to \frac{F_{Q}^{\pi}(-Q^2) [\textrm{\cref{eq:SRinf2}]}}{F_{Q}^{\pi}(s_{th})^{(1- [\textrm{\cref{eq:TheSumRule}}])(\frac{Q^2}{s_{th}})^{3/2}}},
\end{equation}
where the equations referred above correspond to the numerical value 
obtained for the corresponding sum rules. Such values might in general 
differ from the exact theoretical one due to errors and truncation. 
In particular, mild variations of \cref{eq:TheSumRule} will imply 
important deviations for $Q^2\gg s_{th}$, illustrating that special 
attention must be payed to this sum rule in the following. 
Similarly, for asymptotically large $s$ values,
the phase $\delta(s)$ in \cref{eq:SubtThZeroPhase} is easily shown to be divergent 
unless the sum rule is fulfilled.
Finally, special attention must be payed to the kernel in DR2
close to threshold, where it converges as an improper integral. 
This implies an apparent large sensitivity to the form factor 
behavior at threshold. This is however not certainly true as 
long as the sum rule in \cref{eq:TheSumRule} is considered.
While this is irrelevant when analyzing a toy model, a minimal discussion 
is necessary when applying DR2 to real data. This is discussed in 
\cref{app:syst}, in connection to the real data analysis, and will serve to 
assess systematic uncertainties.
Before applying them to real data, where information is available only for a 
finite range of energies, we discriminate the expectations on convergence 
with the help of a toy model. This will be helpful to assess the 
applicability to the available datasets.

\section{Toy model: convergence issues \label{sec:toymodel}}

In the following, and for illustration purposes and in order to gather some
numerical insight, we shall make use of a simplified toy model based on
Ref.~\cite{GomezDumm:2000fz}, that provides with a resummation of
unitary loops at the one-loop level,
\begin{align}
    F_Q^{\pi}(s) &{}= \frac{m_{\rho}^2}{m_{\rho}^2 -s -\frac{192\pi}{\beta_{\rho}^3} \frac{\Gamma_{\rho}}{m_{\rho}}H_{\pi\pi}(s,\mu)}, \\
    H_{\pi\pi}(s,\mu)&{} = \Big\{(s -4m_{\pi}^2)\bar{B}_0(s;m_{\pi},m_{\pi}) \nonumber \\ 
                     & \qquad  - (s/3)\left[1 +3\ln(m_{\pi}^2/\mu^2)\right] \Big\},
\end{align}
\begin{equation}
    \bar{B}_0(s;m_{\pi},m_{\pi}) = 2 +\beta\ln \frac{\beta -1}{\beta +1},
\end{equation}
where $\beta=\beta(s)=\sqrt{1 -4m_{\pi}^2/s}$ and $\beta_{\rho} = \beta(m_{\rho}^2)$. We choose $m_{\rho} = 850$~MeV and $\Gamma = 190$~MeV, that leads to a pole in the second Riemann sheet at $\sqrt{s}=(0.77 -i 0.15/2)$~GeV.\footnote{While this model has a pole in the deep Euclidean region (around $666$~GeV), it serves for our purposes, that focus in the GeV region.} 

\subsection{Convergence I: sum rules \label{sec:testSR}}

To begin with, we comment to which extent the sum rules in \cref{eq:sumrules} are fulfilled for a finite cutoff $s < \Lambda^2$, that would represent a real case where information is only available on a finite energy range. Choosing  $\Lambda=\{1,3,10,100\}$~GeV, we find $\{1.4, 1.3, 1.12, 1.05\}$ for the first sum rule and $\{0.4, 0.3, 0.1, 0.02\}$ for the second one. Both of them display a poor convergence, discouraging its use to search for potential zeroes as done in \cite{Cronstrom:1973wr}.  
We continue with the sum rule in \cref{eq:TheSumRule}. In this respect it is relevant to note that, provided $|F_Q^{\pi}(s)| < F_Q^{\pi}(s_{th})$ for $s>\Lambda_c$, that not only holds in this model but is supported by the experimental data and asymptotics, the lhs. of \cref{eq:TheSumRule} should be approached from above. In this sense, any value below 1 for a finite (but sufficiently large) cutoff would be unacceptable. However, not any positive value would be admissible, for the pQCD prediction (which lies below experimental data) suggests the remainder from $\Lambda=\{1,3,10,100 \}~\textrm{GeV}$ up to infinity to be not more negative than $-\{0.075,0.0054,2\times 10^{-4}, 4\times 10^{-7}\}$, respectively. Turning to the numerics, taking an upper cutoff $\Lambda^2$ we find the values for \cref{eq:TheSumRule} to be $\{ 1.02,1.004,1.0002, 1+3\times10^{-7} \}$, displaying a nice convergence and providing an useful test for any possible parameterization of the form factor.

\subsection{Convergence II: phase and spacelike behavior\label{sec:phaseANDspacelike}}

\begin{figure}[h]
    \centering
\includegraphics[width=0.48\textwidth]{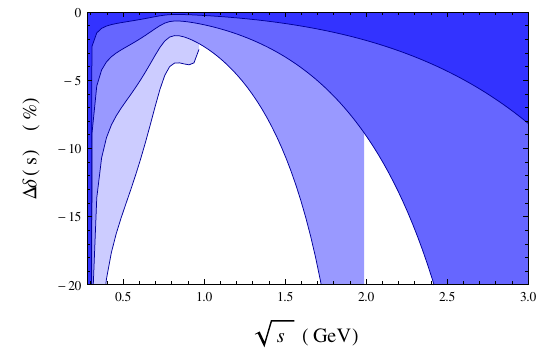} 
\includegraphics[width=0.48\textwidth]{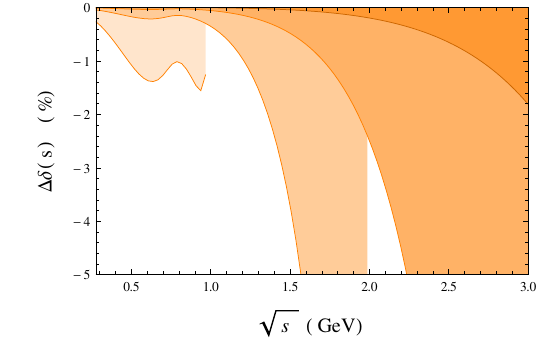}
    \caption{The relative uncertainty for the dispersion relations in \cref{eq:SubtZeroPhase} (top) and \cref{eq:SubtThZeroPhase} (bottom). The phase is never extrapolated beyond the cutoff. From lighter to darker bands the results are shown for $\Lambda=\{1,2,3,5\}$~GeV (see details in the text).}
    \label{fig:ToyModelPhase}
\end{figure}

Next, we explore the rate of convergence for the quantities that we wish to explore with real data. Starting with the phase, we explore the convergence properties of DR1 and DR2. In the following, we compute the phase $\delta_{\Lambda}(s)$ that is obtained if the integral is cut off at $\Lambda=\{1,2,3,5\}$~GeV and plot the relative uncertainty, $\Delta\delta(s) = \delta_{\Lambda}(s)/\delta(s) -1$ in \cref{fig:ToyModelPhase}. As shown, the error is at the few percent level for DR1 when a cutoff $\Lambda=3$~GeV is taken,\footnote{Similar results, albeit with a better performance at threshold, would be obtained for an unsubtracted version of DR2.} albeit with significant uncertainties close to threshold, that reflect the slow convergence of the sum rule in \cref{eq:sumrules} (right). On turn, DR2 greatly improves on convergence (see \cref{fig:ToyModelPhase}) ---and becomes especially relevant at threshold--- representing our preferred choice.

\begin{figure}
    \centering
    \includegraphics[width=0.48\textwidth]{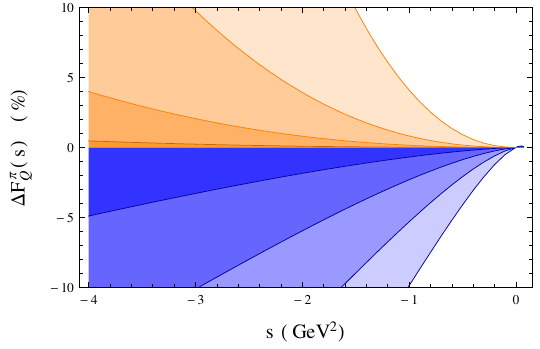}
    \caption{The relative precision in the spacelike region for cutoffs $\Lambda=\{1,2,3,5\}$~GeV shown as blue/orange bands for \cref{eq:SubtZero}/\cref{eq:SubtThZero}.}
    \label{fig:SLtoymodel}	
\end{figure}

We repeat the same exercise, but extrapolating the form factor to the spacelike region. Here, we find a nice property analogous to the sum rule discussed in the previous subsection. Namely, provided that $|F_Q^{\pi}(s)| < F_Q^{\pi}(s_{th})$ for $s>\Lambda_c$, the spacelike values are approached from below and above for DR1 and DR2, respectively. This is interesting, as it allows to set upper and lower bounds in a model-independent way without the need to provide a high-energy completion, which could be interesting in the context of Refs.~\cite{Cheng:2020vwr,Chai:2022ipu}. The results are shown in \cref{fig:SLtoymodel}.

\subsection{Convergence III: the radius and higher derivatives \label{eq:modelradius}}

It is customary to define the coefficients for the form factor series expansion around $s=0$ as
\begin{equation}
    F_Q^{\pi}(s) = 1 +b_{\pi} s + c_{\pi}s^2 + d_{\pi}s^3 + ...
\end{equation}
In particular, the slope parameter has been intensively studied, and it is related to the charge radius via $\langle r_{\pi}^2 \rangle = 6 b_{\pi}$ (see Ref.~\cite{Gonzalez-Solis:2019iod} for an updated list of different estimates). Using DR1 and DR2, the slope reads
\begin{align}
    b_{\pi} &{}= \frac{2m_{\pi}}{\pi}\int_{s_{th}}^{\infty} \frac{\ln|F_Q^{\pi}(z)|}{z^2(z-s_{th})^{1/2}}, \label{eq:SubtZeroRadius}\\
   b_{\pi} &{}=  \frac{3}{2}\frac{\ln F_Q^{\pi}(s_{th})}{s_{th}}
    -\frac{s_{th}^{3/2}}{\pi}\int_{s_{th}}^{\infty} \frac{\ln|F_Q^{\pi}(z)/F_Q^{\pi}(s_{th})|}{z^2(z-s_{th})^{3/2}}, \label{eq:SubtZeroThRadius}
\end{align}
whereas similar expressions can be found for higher derivatives. 
Once more, taking upper cutoffs  $\Lambda=\{1,2,3,5\}$~GeV, we obtain a relative uncertainty of $\{2,1,0.4,0.1\}\%$ for  $b_{\pi}$ when using DR1, and $-\{0.6,0.01,0.002,0.0002\}\%$ when using DR2. For the $n$-th derivative the error scales as $(\Delta b_{\pi})^{2n}$.

\section{Real data analysis\label{sec:realdata}}

With the hindsight of previous section, we analyze the data from Babar
Coll.~\cite{Babar:2012bdw} for the following reasons: first, it is
amongst the most precise extractions, including the finest binning of
2~MeV in the $\sqrt{s}\in(0.5,1)$~GeV region, allowing to check the
validity of the chosen interpolating function; second, it provides
with the largest dataset, ranging from $\sqrt{s}=300$~MeV to
$\sqrt{s}=3$~GeV, thus allowing to compare DR1 and DR2 (we recall that
DR1 requires cutoffs $\Lambda>1$~GeV to provide accurate results) and
to cross-check the sum rule \cref{eq:TheSumRule} that guarantees the
reliability of DR2. Comparatively, other highly-precise datasets are
not in such a privileged situation (see Ref.~\cite{Davier:2023fpl} for
an up-to-date comparison of datasets at the $\rho$ peak). For
instance, the KLOE datasets~\cite{KLOE-2:2017fda}, that are in tension
with BaBar in the common overlapping regimes, have a similar precision
and a binning size of $10$~MeV. However, they only cover the region
from threshold up to $945$~MeV inducing greater systematic
uncertainties and hence largely preventing the current analysis where
systematic uncertainties are below statistical ones. Analyzing KLOE
data on its own would introduce then large systematic uncertainties
for the reasons outlined above. As such, we have restrained ourselves
to the BaBar dataset. Similar comments apply for instance to the
recent results from the CMD-3 collaboration~\cite{CMD-3:2023alj}. We
return to this point later on.

In the following, to {\emph{interpolate}} the data to perform the
numeric integrals we make use of the Gounaris-Sakurai model~\cite{Gounaris:1968mw} used in
Ref.~\cite{Babar:2012bdw}, while modifying the $\rho-\omega$ mixing to
vanish at $s=0$ as argued in \cite{Sanchez-Puertas:2021eqj}, and
including and analogous term for the $\phi$ meson, that is also
visible in the data (see further details in
\cref{sec:phaseshiftOctet}).  Importantly, we subtract corrections
from final state radiation from the data, see
Ref.~\cite{Colangelo:2018mtw}.\footnote{We emphasize that Babar data
  is undressed from intermediate hadronic vacuum polarization (HVP) effects as well, thus free of
  intermediate $1\gamma$ reducible contributions.} We emphasize that,
even if the model is motivated by analyticity and implements the
$\pi\pi$ threshold, there are constant complex phases breaking
unitarity and the Schwarz reflection principle since the form factor
should become purely real below the $\pi^+\pi^-$ threshold. The model should
only be thought of as an interpolator for the experimentally accessible
modulus $|F_Q^{\pi}(s)|$, whereas the physical phase is derived from
\cref{eq:SubtZeroPhase,eq:SubtThZeroPhase}, that respect unitarity and
will differ in general with respect to the original model. 
As we shall see, this has a numerical impact. In summary, the
dispersion relations \cref{eq:SubtZeroPhase,eq:SubtThZeroPhase}
unitarize the phase of the form factor, that represents our main
interest in this work. Further, with the phase at hand, it is possible
to recover the modulus via the Omn{\`e}s solution, which consistency
with the input provides a sanity check. Since \cref{eq:OmnesSubt} has
a slow convergence, we postpone our comments to \cref{sec:radius},
which results allow to set a twice-subtracted version and shows
consistent results. 

\begin{figure}
    \centering
    \includegraphics[width=0.49\textwidth]{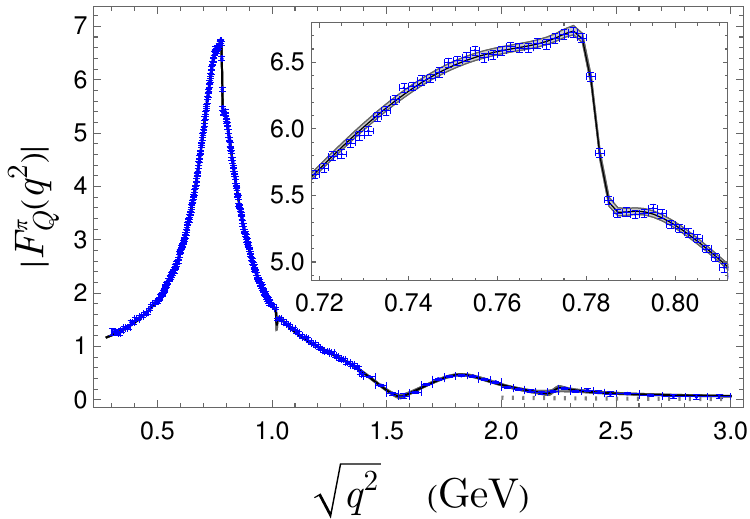}
    \caption{The absolute value of the form factor from the model 
    fitted to Babar data with a $68\%$~CL (gray) band. The data points (blue crosses) 
    correspond to Babar~\cite{Babar:2012bdw}. The zoom shows the $\rho-\omega$ energy 
    region. The dotted gray line represents the pQCD prediction}
    \label{fig:myFitBabar}
\end{figure}

In order to derive uncertainties and to keep track of correlations, we
perform fits to pseudodata (\emph{pseudofits}), that are generated as
replicas of the actual experiment by means of the Monte Carlo method
accounting for the full covariance matrix provided by BaBar
collaboration,\footnote{We found the statistical correlation matrix provided by the BaBar collaboration to
  have near-zero eigenvalues possibly due to the small energy binning, causing numerical difficulties. Such
  results are unstable if considering uncertainties on
  uncertainties. In particular, if rescaling the off-diagonal
  correlation matrix elements by 0.99 such problem is avoided ---an
  approach that we adopt in the following.} including a consistent
treatment of systematic uncertainties in order to avoid d'Agostini
bias~\cite{Ball:2009qv,Colangelo:2018mtw}. Each of the quantities
discussed in the following section is obtained for each
\emph{pseudofit}. The obtained distribution allows then to derive
uncertainty bands at the desired confidence level (CL), fully accounting
for correlations.  Concerning the central fit, corresponding to the most likely parameters, we obtain
$\chi^2/\textrm{dof}=353/317$. Importantly, we find for the sum rule
\cref{eq:TheSumRule} a value of $1.003$ taking a cutoff of
$\Lambda=3$~GeV, in line with our model expectations and ensuring
reliable results for DR2.\footnote{Interesting enough, if fitting to
  the exponential parameterization in Ref.~\cite{Guerrero:1997ku},
  values below $1$ are obtained, pointing to a violation of the sum
  rule and unreliable results. Indeed, we checked that, within such a
  model, the results from DR1 and DR2 are not equivalent. On turn, if
  we use the model in \cref{sec:toymodel} instead, the sum rule is
  satisfied and the results are nearly indistinguishable,
  that emphasizes once more the relevance of the sum rule.}  
Concerning pseudofits, most of them display reasonable
results for the sum rule. Still, in order to avoid physically
unacceptable outcomes, those pseudofits in which the sum rule result
is either negative or above 1.0054 are discarded for the reasons
outlined in \cref{sec:testSR}, keeping only the ones surpassing this
test to derive the physical quantities in the following.  In
\cref{fig:myFitBabar} we show the absolute value of the form factor
with 68\%CL bands. We emphasize that our bands are similar to the data
uncertainties close to the $\rho$ peak. Finally, we also provide the
value for the form factor at threshold, $F_{Q}^{\pi}(4m_{\pi}^2) =
1.174(1)$, that is ubiquitous when using DR2. Our value is in nice
agreement with the model-independent prediction $1.176(2)$ from
Ref.~\cite{Ananthanarayan:2018nyx}. In the following subsections, we
discuss the different outcomes, starting with the phase of the form
factor, that represents our main result.

\subsection{The phase of the charged form factor\label{sec:FFphase}}

\begin{figure}
    \centering
    \includegraphics[width=0.485\textwidth]{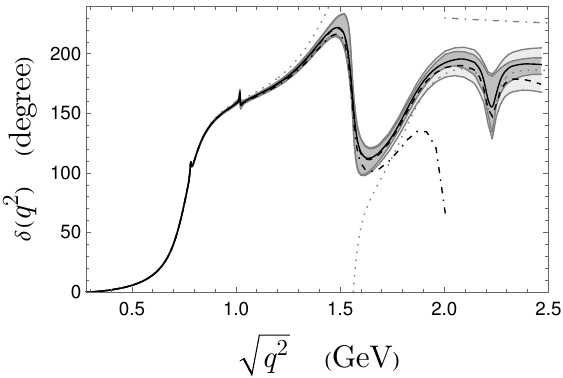} \\ \vspace{2mm}
    \includegraphics[width=0.485\textwidth]{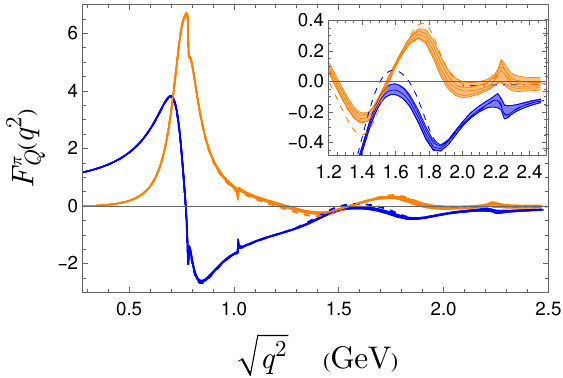}
    \caption{Top: The phase of the form factor from DR2 with an upper cutoff 
    $\Lambda=5$~GeV (black solid line with $68\%$~CL gray band; the outer band includes 
    systematics from interpolation). 
    The value obtained for $\Lambda=2/3$~GeV is shown as a 
    dot-dashed/dashed black line, and the original phase from the 
    model as a gray-dotted line. The pQCD prediction is shown as a dot-dashed gray 
    line close to $225^{\circ}$.  
    Bottom: The real and imaginary parts obtained from DR2 are shown in blue 
    and orange, respectively, with corresponding $68\%$~CL bands; the outer band includes 
    systematics.
    The real(imaginary) part from the model is shown as a blue(orange)-dashed 
    line.}
    \label{fig:PhaseReIm}
\end{figure}

In this section, we extract the phase of the form factor, which is the main object of interest in this work. While this is closely related to the $\delta_1^1$ phase-shift, comparison requires the removal of the isospin zero component of the electromagnetic form factor, that we postpone till next section. Together with the modulus, the phase allows to extract the real and imaginary parts.
Our result for the phase of the form factor obtained from DR2\footnote{We checked that the phase obtained through DR1 is very similar, with tiny differences, especially near threshold and at high energies, expected from the convergence pattern discussed in \cref{sec:toymodel}, that serves as a cross-check.} is shown in \cref{fig:PhaseReIm} (top) for a cutoff $\Lambda=5$~GeV. Such a large cutoff is unnecessary below $\sqrt{s}\simeq 1.8$~GeV, but required above. Regarding the potential systematics of such an extrapolation, we note that the dashed line result (from choosing $\Lambda=3$~GeV) provides a lower bound. An extreme upper bound would be obtained if extrapolating the model with pQCD above, that is however a factor of four smaller at 3~GeV and likely unrealistic and leads to $\delta(2.5~\textrm{GeV}^{2}) =201^{\circ}$. It is hard to imagine such a sudden drop immediately after 3~GeV. Indeed, would we extrapolate our model just up to 3.5/4~GeV, we would find $\delta(2.5~\textrm{GeV}^{2}) =(184/188)^{\circ}$, pretty close to our central value of $190.5^{\circ}$. Other approaches including $F_Q^{\pi}(s) \to F_Q^{\pi}(\Lambda^2)\Lambda^2 (\alpha_s(s)) /s $ lead to $191.5(191.8)^{\circ}$. Again, in line with our central result. Likewise, our plot shows systematic uncertainties from extrapolation at threshold, that we discuss in \cref{app:syst}. In that plot, we also show the phase that would be obtained directly from the model as a dotted-gray line. This needs not agree (and indeed does not agree) with the one obtained through DR2. It especially displays marked differences close to the $\rho'$. Indeed, the phase that is obtained directly from the model approaches $3\pi$ at infinity, which is only possible in the presence of (complex-conjugate); such possibility and its potential impact is discussed in detail in \cref{app:zeros}. 
Note however that: (i) the complex phases in the original (interpolating Gounaris-Sakurai) model violate the Schwarz reflection principle and unitarity, that requires them to be dynamical; (ii) the resonance description is well oversimplified above the $\rho$ region, where many inelastic channels open. Overall, such considerations cast serious doubts on the validity of the original (interpolating Gounaris-Sakurai) model's phase and its extrapolation to the complex plane.
By contrast, in the absence of zeros, the current approach cures such pathologies and diminishes potential model-dependencies by unitarizing the phase of the model and providing an extension to the complex plane consistent with analyticity and unitarity constraints. The differences can also be appreciated when plotting the real and imaginary parts of the form factor, see \cref{fig:PhaseReIm} (bottom). 
These results are very interesting since, in the absence of zeros, they allow to extract the phase above inelastic thresholds, where $\delta(s)$ no longer identifies with $\delta_1^1(s)$, directly from data with reduced model dependencies. Note in this respect that current estimates are mostly based on models that violate analyticity~\cite{Roig:2011iv,Schneider:2012ez,Gonzalez-Solis:2019iod}. 
In the following section, we derive the relation to the isovector $\delta_1^1$ phase-shift and clarify common misunderstandings around this.

\subsection{The isovector phase-shift and the octet form factor\label{sec:phaseshiftOctet}}

As argued in Refs.~\cite{Sanchez-Puertas:2021eqj,Broniowski:2021awb}, the electromagnetic form factor is not a pure isovector $I=1$ object, but receives non-vanishing IB $(I=0)$ contributions. In particular, the electromagnetic current can be decomposed as $J^{\mu}_{Q} = V^{\mu}_3 + (1/\sqrt{3})V^{\mu}_8$, where $V^{\mu}_a = (1/2)\bar{q}\gamma^{\mu}\lambda^a q$ are the usual $SU(3)_V$ currents, with $\lambda^a$ Gell-Mann matrices. This implies
\begin{multline}
    \bra{\pi^+(p')} J_Q^{\mu} \ket{\pi^+(p)} =
    \bra{\pi^+(p')} V_3^{\mu} +\frac{1}{\sqrt{3}}V_8^{\mu} \ket{\pi^+(p)} \\ = \left(F_3^{\pi}(q^2) +\frac{1}{\sqrt{3}}F_8^{\pi}(q^2) \right)(p+p')^{\mu} .   
\end{multline}
The octet part would vanish in the isospin-symmetric limit upon $G$-parity, while IB effects drive a nonzero form factor. Still, the null octet charge of the pion demands $F_8^{\pi}(0)=0$. Note that, in the $SU(2)$ limit, the octet part reduces to the baryonic one discussed in \cite{Sanchez-Puertas:2021eqj,Broniowski:2021awb}. Indeed, these would be identical (up to overall constants), in the large-$N_c$ limit of QCD, since strange-quark effects find a further suppression following from the OZI rule. 
With this decomposition in mind, we discuss the resonances that should appear in each form factor following Ref.~\cite{Sanchez-Puertas:2021eqj}. In particular, regarding the isovector form factor, $F_3^{\pi}(q^2)$, the $\rho$ vector meson (alternatively, $\pi^+\pi^-$ rescattering) appears at $\mathcal{O}(0)$ in IB, whereas the $\omega, \phi$ resonances (say $3\pi,K\bar{K}$ intermediate effects) appear at $\mathcal{O}(2)$. Indeed, the leading IB effect in the electromagnetic form factor comes from the octet form factor, $F_8^{\pi}(q^2)$, where intermediate $\rho$ and $\omega$ states appear both at $\mathcal{O}(1)$ in IB; the $\phi$ resonance is, in addition, OZI suppressed and is expected to be subleading, yet visible in the data. This means that the observed effects of the $\omega$, $\phi$ resonances are a feature of the octet form factor, whereas such effects play a marginal role in the isovector one, which phase would correspond to $\delta_1^1$. Indeed the latter quantities receives their leading IB effects at $\mathcal{O}(2)$. This contrasts with statements in Refs.~\cite{Benayoun:2019zwh,Benayoun:2023dkl} that compare the phase of the electromagnetic form factor to $\delta_1^1$, which is inconsistent.\footnote{We note in addition that intermediate photonic states should be ignored here, since the Babar cross section is removed from HVP effects. The leading electromagnetic IB effects would come from intermediate $h\gamma$ states, see for instance Ref.~\cite{Colangelo:2022prz}.} 

As such, we need to disentangle the isovector and octet form factors in order to extract the $\delta_1^1$ phase-shift, that requires a minimal modelling. To do so, in the previous fitting procedure we have employed a similar model to that in Ref.~\cite{Sanchez-Puertas:2021eqj} to interpolate the data, but including the $\phi$ meson. Specifically,
\begin{multline}\label{eq:GSpar}
        F_Q^{\pi} = \Big( D_{\rho}(s)[1 +c_{\omega} s D_{\omega}(s)  +c_{\phi} s D_{\phi}(s)] 
    +c_{\rho'}D_{\rho'}(s) \\ +c_{\rho''}D_{\rho''}(s) +c_{\rho'''}D_{\rho'''}(s) \Big)
    \frac{1}{1 +c_{\rho'} +c_{\rho''} +c_{\rho'''}} ,
\end{multline}
where $c_X$ are complex parameters, $D_{\rho}(s)$ is the Gounaris-Sakurai parameterization in Ref.~\cite{Babar:2012bdw}  and $D_{\omega,\phi}$ are modelled through a normalized Breit-Wigner parameterization as in Ref.~\cite{Babar:2012bdw}. In addition, while the $\omega$ parameters are obtained from the experiment, the $\phi$ mass and width need to be fixed to the PDG values. To recover the purely isovector part, we set $c_{\omega,\phi}\to0$, where little model-dependence is expected. This allows to obtain the absolute value of the isovector form factor and to recover the phase through DR2, along the lines of previous section. In particular, we find $c_{\omega} = 0.00196(6) e^{i 0.06(5)}~\textrm{GeV}^{-2}$ and $c_{\phi} = 0.0008(3) e^{i 1(1)}~\textrm{GeV}^{-2}$, displaying a precise result for the $\omega$ contribution, in contrast to the $\phi$ case. A better assessment of the role of the $\phi$ meson might be obtained in the future using CMD-3 data~\cite{CMD-3:2023alj} that, however, requires a dedicated analysis due to the tensions with respect to Babar data here employed. 
\begin{figure*}
    \centering
    \includegraphics[width=.49\textwidth]{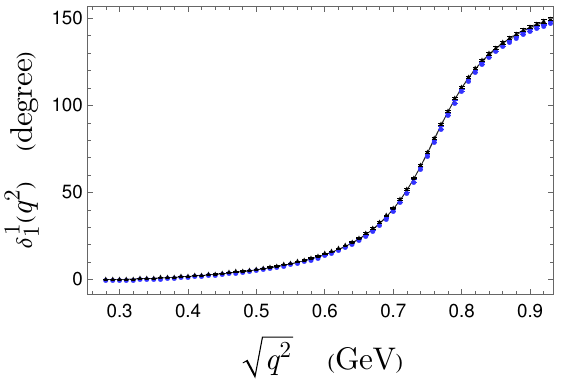}
    \includegraphics[width=.49\textwidth]{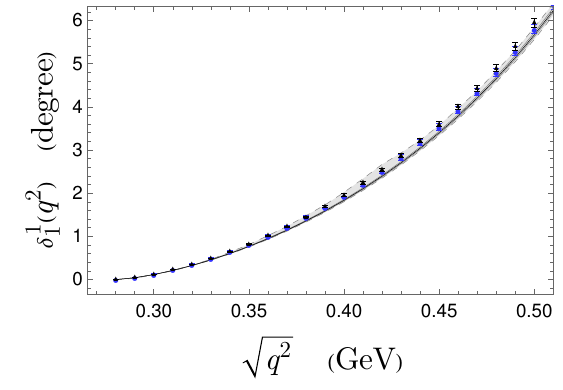} \\
    \includegraphics[width=.49\textwidth]{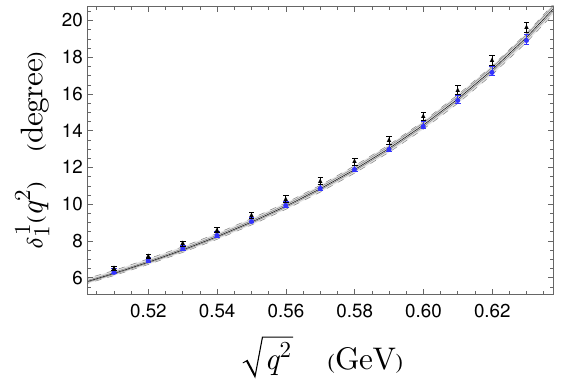}
    \includegraphics[width=.49\textwidth]{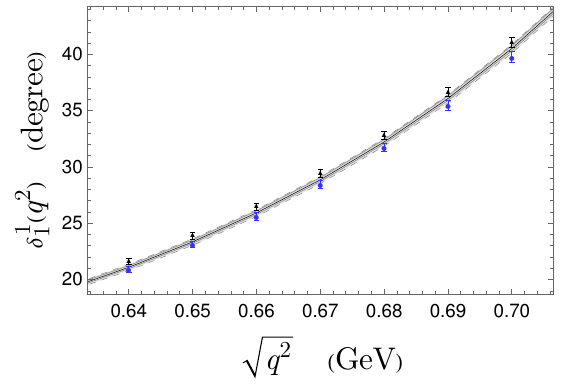} \\
    \includegraphics[width=.49\textwidth]{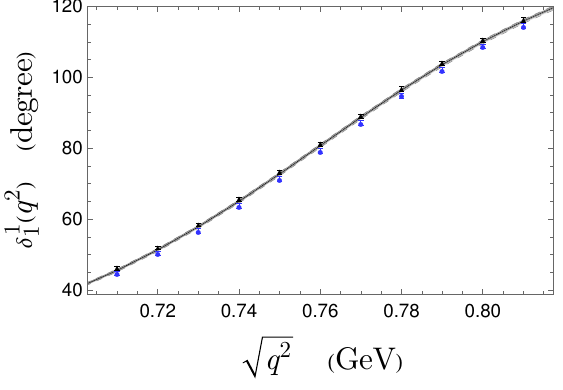}
    \includegraphics[width=.49\textwidth]{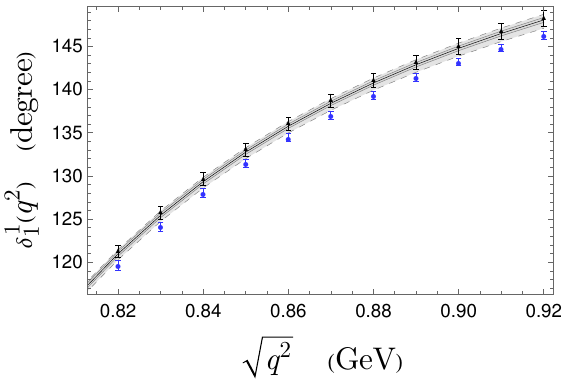}
    \caption{We show our extraction of $\delta_1^1(s)$ with $68\%$~CL (gray) bands; the additional lighter-gray band includes systematic uncertainties. 
    We confront our results against the Madrid~\cite{Garcia-Martin:2011iqs} (blue dots) and Bern~\cite{Colangelo:2018mtw} (black triangles) phases.}
    \label{fig:PhaseShift} 
\end{figure*}
With these results, the isovector phase-shift is shown in \cref{fig:PhaseShift}. 
This compares to the $\delta_1^1(s)$ phase-shift below the first relevant 
inelasticity opening at $\sqrt{s}=m_{\omega} +m_{\pi^0}$. 
The plot shows the comparison to the Madrid~\cite{Garcia-Martin:2011iqs} 
and Bern~\cite{Colangelo:2018mtw,Colangelo:2020lcg} phases 
(which are, overall, not overlapping). 
At this level of precision, the systematics stemming from the extrapolation 
down to threshold become relevant. 
These are discussed in \cref{app:syst} and are notoriously constrained 
by the sum rule. Still, they are especially relevant at low energies, 
where these dominate and cannot compete in precision with 
Refs.~\cite{Garcia-Martin:2011iqs,Colangelo:2018mtw,Colangelo:2020lcg}. 
Additional systematics from the impact of 
potential zeros require a separate study and are discussed in \cref{app:zeros}.
The corresponding error band is shown in \cref{fig:PhaseShift}. 
Overall, we observe a good agreement with both phases, with an average deviation of about $2\sigma$ for both cases that reduces below the 1$\sigma$ level when accounting for systematics. Noteworthy, our results align better at high-energies with the Bern phase, that features an up-to-date analysis of $e^+e^-\to\pi^+\pi^-$ data. This is reflected in their value at the matching point, $\delta_1^1(0.8~\textrm{GeV})=110.4(7)^{\circ}$, which is in good agreement with ours, $110.1(3)^{\circ}$, yet $3\sigma$ away from the Madrid result, $108.6(6)^{\circ}$.
To put our results on a firmer ground, it would be interesting to take advantage of the densely populated datasets to evaluate the integrals numerically without resorting to an interpolation method. While this seems plausible above $\sqrt{s}= 500$~MeV (see \cref{app:syst}), it would require a dedicated effort in the region below, which goes beyond the scope of this work. 

\begin{figure}
    \centering
    \includegraphics[width=.49\textwidth]{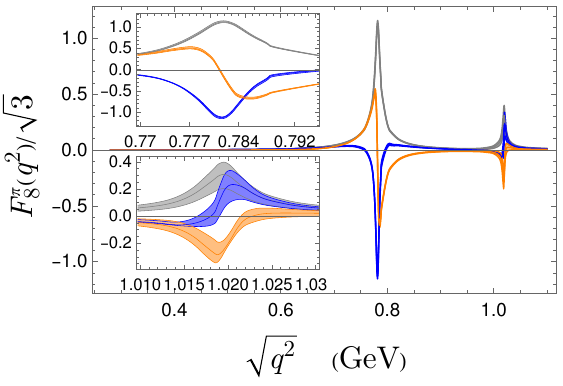}\\
    \includegraphics[width=.49\textwidth]{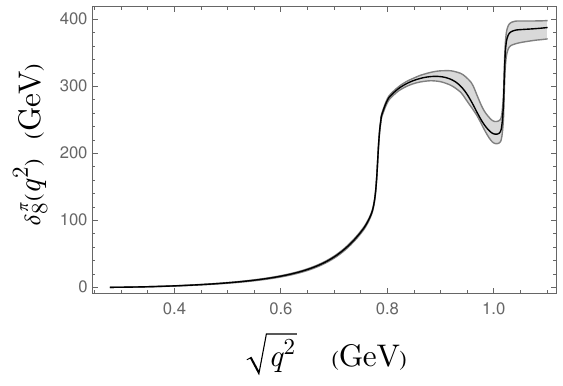}
    \caption{Top: our result for $F_8^{\pi}(q^2)$ displaying the absolute value (gray) as well as the real (blue) and imaginary (orange) parts (bands stand for 68\% CL). Bottom: the phase of the octet form factor (black) with 68\% CL gray band.}
    \label{fig:octetfigs}
\end{figure}

The results from \cref{sec:FFphase} for the real and imaginary parts of the form factor can be used together with previous results to subtract the isovector form factor and to obtain the octet form factor.\footnote{In practice, it is simpler to use $\delta_Q$ and the phase difference, $\delta_Q-\delta_3$, which is less noisy.} We show our results in \cref{fig:octetfigs}. We emphasize that our results are not reliable much above 1~GeV, since our model lacks the $\rho',\omega'$ contributions, that cannot be obtained from data in contrast to the narrow $\omega,\phi$ resonances. 
Compared to the isovector case, we find a vanishing value for the form factor at zero momentum transfer, and a sudden change in sign for the imaginary part due to the $\omega$ resonance at $\sqrt{s} \sim m_\omega$. This is also reflected in the phase, that rapidly increases by $\pi$ when crossing the $\omega$ resonance. The $\phi$ meson effects are clearly visible but, compared to their $\rho,\omega$ counterparts, suffer from larger uncertainties. A distinct feature in this form factor when compared to the isovector one regards its phase, that seems to approach 2$\pi$ asymptotically. This makes perfect sense from the point of view of the Omnès-like reconstruction, that would be possible for $\ln F_8^{\pi}(s)/s$, and would demand a factor of $s$ in front of \cref{eq:OmnesSubt}, as well as with the argument theorem, since the zero at $s=0$ demands an extra factor of $\pi$. Overall, that would imply a $s^{-1}$ asymptotic behavior, in accordance with pQCD.\footnote{Note that, for the octet form factor, the pQCD behavior is largely unknown, as it would depend on the size of the leading odd Gegenbauer polynomial (odd polynomials reflect asymmetries in $u/d$ quark distributions and vanish in the IB limit), which is unknown. Still, pQCD would imply $s^{-1}$ behavior modulo $\alpha_s$ corrections.} We emphasize that such phase no longer identifies with a scattering phase, but would still relate, upon unitarity, to the $\delta_1^1$ phase and the $P$-wave $\braket{\pi^+\pi^-\pi^0 | \pi^+\pi^-}$ phase-shift. Systematic uncertainties are not included in the quoted plots, while their size would be similar to previous cases.

This completes our discussion on phases. In the following, we discuss the information that can be obtained for the radius and higher derivatives of the electromagnetic form factor.

\subsection{The pion radius and their companions\label{sec:radius}}

Next, we move on to the pion radius. Using the DR in \cref{eq:SubtZeroThRadius} with a cutoff $\Lambda=3$~GeV we obtain 
\begin{multline}
    \langle r_{\pi}^2 \rangle =  11.01(7)_{\textrm{st}}(^{+{10}}_{-4})_{\textrm{sys}}\textrm{GeV}^{-2} = 
       \\ = (0.655(2)_{\textrm{st}}(^{+4}_{-2})_{\textrm{sys}}~\textrm{fm})^2 = 0.429(2)_{\textrm{st}}(^{+3}_{-1})_{\textrm{sys}}~\textrm{fm}^2 ,
\end{multline}
Our result is in good agreement with the model-independent estimate from Ref.~\cite{Ananthanarayan:2017efc}, $\langle r_{\pi}^2 \rangle = (0.657(3))~\textrm{fm}^2$ and with the recent estimate from Ref.~\cite{Colangelo:2018mtw} based on a dispersive representation, $\langle r_{\pi}^2 \rangle = 0.429(4)~\textrm{fm}^2$, albeit smaller than $11.28(8)\textrm{GeV}^{-2}$ from  Ref.~\cite{Gonzalez-Solis:2019iod} based on a dispersive fit to $\tau$ data from Belle collaboration (see Table~6 in Ref.~\cite{Gonzalez-Solis:2019iod} for a detailed compilation of different estimates). As a check of consistency, we obtain $\langle r_{\pi}^2 \rangle = (0.656(2)~\textrm{fm})^2$ using \cref{eq:SubtZeroRadius}, in agreement with expectations from \cref{sec:toymodel}. 

As a nice byproduct, with the knowledge of the radius at hand, it is possible to reliably reconstruct the absolute value of the form factor using a twice-subtracted Omn{\`e}s-like dispersion relation with a cutoff $\Lambda=3$~GeV, that should recover the input value for the modulus of the form factor. The method proves to be self-consistent within uncertainties, reinforcing the reliability of our results. 

Finally, we compute for completeness the quadratic and cubic slope, finding
\begin{align}
    c_{\pi} =&{} 3.84(3)_{\textrm{st}}(^{+5}_{-2})_{\textrm{sys}}~\textrm{GeV}^{-4}, \\
    d_{\pi} =&{} 10.1(1)_{\textrm{st}}(^{+3}_{-1})_{\textrm{sys}}~\textrm{GeV}^{-6},
\end{align}
with systematic uncertainties arising from interpolation, see \cref{app:syst}. These are in good agreement with the bounds from \cite{Ananthanarayan:2013dpa}, $c_{\pi} \in (3.79,4)~\textrm{GeV}^{-4}$ and $d_{\pi} \in (10.14,10.56)~\textrm{GeV}^{-6}$, albeit smaller than  $c_{\pi} = 3.94(4)\textrm{GeV}^{-4}$ and $d_{\pi} = 10.54(5)\textrm{GeV}^{-6}$ from Ref.~\cite{Gonzalez-Solis:2019iod} (find a complete compilation in \cite{Gonzalez-Solis:2019iod}). Note in this respect that results in \cite{Gonzalez-Solis:2019iod} come from $\tau$ data, that need not agree with this case since comparison requires to correct the data, including IB effects, see for instance Refs.~\cite{Cirigliano:2002pv,Davier:2010fmf,Masjuan:2023qsp}. Furthermore, there is no consensus on the agreement of $\tau$ and $e^+e^-$ data after IB corrections are accounted for. In the following section, we discuss the bounds that can be derived in the SL region.

\subsection{Extrapolation to the spacelike region \label{sec:dataSL}}

\begin{figure}
    \centering
    \includegraphics[width=0.49\textwidth]{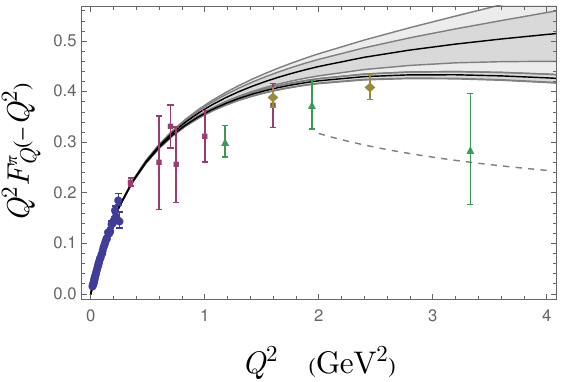}
    \caption{The extrapolation to the spacelike region using \cref{eq:SubtZero} (lower gray band) and \cref{eq:SubtThZero} (upper gray band). They provide an estimate of the lower and upper values for the form factor. We also provide lighter error bands corresponding to combined statistic and systematic uncertainties. In addition, we provide the NLO pQCD prediction (dashed-gray line). Experimental results from \cite{Bebek:1974ww,Bebek:1977pe} (green triangles), \cite{NA7:1986vav} (blue dots), \cite{JeffersonLabFpi-2:2006ysh} (orange diamonds) and \cite{JeffersonLabFpi:2007vir} (purple squares; includes compilation therein).
    }
    \label{fig:SLresult}
\end{figure}

As argued in \cref{sec:phaseANDspacelike}, under the mentioned assumptions, DR1 and DR2 provide lower and upper bounds for the form factor in the spacelike region, while we expect those from DR2 to lie closer to the real value. We show our results using an upper cutoff $\Lambda=3$~GeV in \cref{fig:SLresult}.
We also check that, if extrapolating to larger cutoffs, both results
come close, providing yet another consistency check. 
Once more, we include, in addition to statistical uncertainties, systematic ones form the extrapolation at threshold. 
These affect more DR2, as it is more sensitive to threshold, while DR1 suffers from larger finite cutoff uncertainties. 
Increasing the cutoff with a model, would improve both, DR1 and DR2.
Our results are
in good agreement with the low-energy data from
\cite{Bebek:1974ww,Bebek:1977pe,Amendolia:1983di,NA7:1986vav,JeffersonLabFpi:2000nlc,JeffersonLabFpi-2:2006ysh,JeffersonLabFpi:2007vir} 
and suggest slightly larger values at larger $Q^2$. In this respect, once more, the data suggests that the pQCD behavior, which prediction is shown as a dashed-gray line in \cref{fig:SLresult}, is not reached below $(2~\textrm{GeV})^2$, which is also in line with the findings in Ref.~\cite{Simula:2023ujs} using the dispersive matrix approach with spacelike data and lattice QCD results. In this respect, it would be interesting to check implications from other experiments, such as KLOE or CMD-3 in order to elucidate whether spacelike data could help assessing which of the currently conflicting timelike datasets could be potentially correct. This requires a better understanding on how to combine the  conflicting datasets and represents work in progress.

\section{Conclusions\label{sec:conclusions}}

In the absence of zeroes it is possible to extract the phase of the pion form factor from its modulus along the cut using dispersion relations. Compared to common dispersive analysis, building on the phase, the input to the dispersion relation is a measurable quantity. In this article, we have revisited this kind of dispersive analysis motivated by the large dataset from Babar Collaboration in the $(2m_{\pi} < \sqrt{s} < 3)$~GeV region and incorporating further subtractions that improve on convergence with respect to previous studies. 
We have made use of an auxiliary (unitarity-breaking) Gounaris-Sakurai fit to interpolate the data and evaluate the DR integral. 
The obtained result proves self-recursive and fulfills the relevant sum rule, thus fulfilling analytic constraints and providing a sanity check of our results.
Furthermore, we have evaluated systematic uncertainties from the interpolation method and the  impact of potential zeros of the form factor. 
In consequence, we have obtained the phase of the form factor below $2.5$~GeV. The fit also models the relevant isovector and octet form factors that appear in the electromagnetic current decomposition. The latter is an isospin-breaking effect and relates to the $\omega,\phi$ interferences, clearly visible in the data. This way we can extract the isovector form factor and corresponding phase-shift, that identifies with the $\delta_1^1$ $\pi\pi$ phase shift below inelasticities, finding a good agreement with respect to the analysis based on Roy equations and competitive uncertainties except for the region close to threshold, which uncertainty is mainly driven by systematics from interpolation. In addition, we have studied the potential effect of zeroes of the form factor. Our method provides however a unique capability to extract the phase above inelasticities, which is currently poorly known and more model dependent. In addition, we have extracted the isovector and isoscalar components of the electromagnetic form factor, 
as well as the pion radius and higher derivatives, which are in good agreement with current determinations. Furthermore, we have set upper and lower bounds on the spacelike behavior without the necessity to model the high-energy part, that otherwise would be prone to model dependencies, including the everlasting debate on the onset of pQCD. Our analysis suggests slightly larger values than current experimental data in the spacelike region. In this respect, it would be interesting to check in the future the implications from other conflicting datasets. Since the latter do not encompass such a large energy range, some kind of combination with Babar is necessary, that represents work in progress. Once this is achieved for CMD-3 data, that has a better coverage of the $\phi$ region, a better extraction of the octet form factor would be achieved, that could be compared with expectations from models. We note this has attracted attention in the context of isospin-breaking corrections in the muon $(g-2)$~\cite{Colangelo:2022prz}.

Finally, the current study suggests interesting lines of further research, such as the possibility to avoid interpolating functions thanks to the dense dataset, or to use the knowledge of the form factor along the cut to extract resonance poles. 

For convenience, we provide our results for the phase and modulus of the form factor in \cref{app:datasets}.\\

\textbf{Acknowledgements:} We acknowledge B.~Malaescu for comments on the Babar covariance matrix. 
We also acknowledge B. Kubis for comments on the draft and M.~Hoferichter for providing the phase from Ref.~\cite{Colangelo:2018mtw}.  
This work has been supported by the European Union’s Horizon 2020 Research and Innovation Program under Grants No. 754510 (EU,
H2020-MSCA-COFUND2016), the Spanish Ministry of Science and Innovation under Grants No. PID2020–114767GB-I00, Junta de Andalucía under Grants No. POSTDOC\_21\_00136 and No. FQM-225.

\appendix

\section{The argument theorem\label{app:argumentTH}}

Given a function $f(z)$ with zeros and poles, 
its logarithmic derivative, $f'(z)/f(z)$ will features simple poles with 
residue $1(-1)$ at the location of the zeros(poles) of the original 
function $f(z)$.
Then, it can be shown that, inside a closed contour where the function 
$f(z)$ is analytic except for the presence $P$ poles, 
\begin{equation}
  \frac{1}{2\pi i} \oint\limits_{C} \frac{f'(z)}{f(z)} dz = N-P,
\end{equation}
with $N$ the number of zeros enclosed inside such contour~\cite{Schaum}. 
This theorem can be easily applied to our function by deforming 
appropriately a closed contour so that it excludes the branch 
cut for $F_{Q}^{\pi}(z)$. In such a way, and choosing a circle with radius $|z|=\Lambda^2$, one obtains
\begin{equation}
  \oint\limits_{C} \frac{F_{Q}^{\pi\prime}(z)}{F_{Q}^{\pi}(z)} dz = 2\pi i(N-P).
\end{equation} 
Regarding the left-hand side of the equation above, the circle at 
threshold provides a vanishing contribution; 
the paths along the branch cut lead to the discontinuity 
$2i\delta(\Lambda^2)$; finally, we evaluate the outer circle contour with pQCD, leading to 
$-2\pi i(1 +L^{-1} +\frac{6.58}{\pi} \frac{4\pi}{\beta_0} L^{-2} +... )$.
Assembling all these quantities, one derives the relation in \cref{eq:argumentP}.
For additional discussions on the interrelation among the phase and the 
zeroes of form factors, see also Ref.~\cite{Ananthanarayan:2004xy}.

\section{Systematic errors from the presence of zeros\label{app:zeros}}

If the form factor features zeroes in the first Riemann sheet, the  phase and modulus dispersion relations must be modified. 
Following Ref.~\cite{Cronstrom:1973wr}, this  amounts to the replacement in Cauchy's representation
$\{1, 1/\sqrt{s_{th} -s} \}\ln F_Q^{\pi}(s) \to \{1, 1/\sqrt{s_{th} -s} \}\ln [F_Q^{\pi}(s)/B(s)]$, where $B(s)$ is a product of conformal factors
\begin{equation}
  B(s) = \prod_i \frac{ \sqrt{s_{th} -z_i}  -  \sqrt{s_{th}-s} }{ \sqrt{s_{th} -z_i}  +  \sqrt{s_{th}-s} } \, ,
\end{equation}
with $z_i$ the zeroes of the form factor. Note that, for the Schwarz reflection principle to hold, complex poles must appear in (complex-conjugate) pairs.  
For the unsubtracted modulus dispersion relation, \cref{eq:DR0}, this implies
\begin{align}\label{eq:DR0}
    F_Q^{\pi}(s) ={}& B(s) 
    \operatorname{exp} \left(
    \frac{\sqrt{s_{th} -s}}{\pi} \! \int_{s_{th}}^{\infty} \!
    \frac{\ln|F_Q^{\pi}(z)| ~dz }{\sqrt{z -s_{th}}(z-s)}
    \right), \\
    \delta(s) ={}& \phi(s)
    -\frac{\sqrt{s -s_{th}}}{\pi}  \operatorname{PV}\!\!\int_{s_{th}}^{\infty} \frac{\ln|F_Q^{\pi}(z)|~dz}{(z-s)\sqrt{z -s_{th}}}, 
\end{align}
where $\phi(s) = \operatorname{arg}B(s)$. As a consequence, the form factor is modified below/above threshold by the value/phase 
of $B(s)$. 
Similarly, the equivalent of DR1 reads (see also Ref.~\cite{Geshkenbein:1998gu})
\begin{align}
    F_{Q}^{\pi}(s) &{}= \frac{B(s)}{ B(0)^{ \sqrt{\frac{s_{th} -s}{s_{th}}} } }\operatorname{exp}\Big(
    \frac{s\sqrt{s_{th} -s}}{\pi} \nonumber \\ & \qquad\qquad\quad \times \int_{s_{th}}^{\infty}
    \frac{\ln|F_Q^{\pi}(z)|~dz }{z\sqrt{z -s_{th}}(z-s)}
    \Big), \label{eq:SubtZero} \\
    \delta(s) &{}= \phi(s) + \sqrt{\frac{s -s_{th}}{s_{th}}} \ln B(0) \nonumber\\ & \quad
    -\frac{s\sqrt{s -s_{th}}}{\pi} \operatorname{PV}\int_{s_{th}}^{\infty}dz 
    \frac{\ln|F_Q^{\pi}(z)|}{z\sqrt{z -s_{th}}(z-s)}, \label{eq:DR1zeros}
\end{align}
and similar considerations follow. 
Similarly, the extraction of the slope parameter will be affected and \cref{eq:SubtZeroRadius} reads now
\begin{equation}
    b_{\pi} = \frac{B'(0)}{B(0)} +\frac{\ln B(0)}{2s_{th}} +\frac{2m_{\pi}}{\pi}\int_{s_{th}}^{\infty} \frac{\ln|F_Q^{\pi}(z)|}{z^2(z-s_{th})^{1/2}}. \\
\end{equation}
In parallel to \cref{eq:sumrules}, one could derive a sum rule from the value at zero. 
However, it features the same convergence properties as \cref{eq:sumrules}.
In brief we will introduce the analogous version of \cref{eq:TheSumRule}, 
that will prove more useful. 
Before that, it makes sense to question whether the presence of zeros would 
forbid a self-recursive function, as we found. This is, whether feeding the phase DR with 
our phase ---obtained under the assumption of no zeros--- would lead back to 
the original modulus from which the phase was extracted. 
To do so, we introduce the phase DR in the presence of zeros, that reads 
in its once- and twice-subtracted forms
\begin{align}
  F_{Q}^{\pi}(s) &{}= \frac{B(s)}{B(0)}\operatorname{exp} \left( \frac{s}{\pi}\int_{s_{th}}^{\infty} dz \frac{\delta(z) -\phi(z)}{z(z-s)}  \right) ,\\
  F_{Q}^{\pi}(s) &{}= \frac{B(s)}{B(0)}\operatorname{exp} \left( s~F_{Q}^{\pi\prime}(0) - s\frac{B'(0)}{B(0)} \right.  \nonumber 
  \\ & \qquad\qquad\quad \left. +\frac{s^2}{\pi}\int_{s_{th}}^{\infty} dz \frac{\delta(z) -\phi(z)}{z^2(z-s)} \right) , \label{eq:DRphase2S0}
\end{align}
respectively. 
Note that, would the absence of zeros be false, we would obtain a shift in $\bar{\delta}(s) = \delta(s) -\phi(s) \ (-\ln B(0)\sqrt{s/s_{th} -1})$. 
Such a shifted phase would be the input of our phase DR. For the twice-subtracted phase DR, it is simple to prove that the shift in DR1 precisely 
reproduces the additional terms in \cref{eq:DRphase2S0}, guaranteeing that the original modulus is recovered. Hence,  such a property cannot 
be taken as a signal of the absence of zeros. 
Note in addition that approaches based on the phase DR (see for instance Ref.~\cite{Gonzalez-Solis:2019iod}), 
where the phase is taken as an input, implicitly assume the absence of zeros and are 
affected by similar considerations, see for instance Ref.~\cite{Bernard:2009zm} for similar discussions 
in the context of the $K\pi$ scalar form factor.

Finally, we turn back to the DR subtracted at threshold which, in the presence of zeros, requires further considerations. In particular, near $s=s_{th}$,
\begin{multline}
g(s) \equiv \frac{1}{(s_{th} -s)^{3/2}} \ln\frac{ F_Q^{\pi}(s) }{F(s_{th})B(s)} \\ 
   \simeq  \frac{-2\sum_i \frac{1}{\sqrt{s_{th} -z_i}} }{s -s_{th}} \equiv \frac{ \operatorname{Res}g(s_{th}) }{ s -s_{th} }
\end{multline}
the function $g(s)$ has a pole. Consequently, the contour chosen for Cauchy's representation must be modified to encircle the pole at threshold, leading to 
\begin{align}
   F_{Q}^{\pi}(s) &{}= F_{Q}^{\pi}(s_{th})  B(s) \operatorname{exp}\Bigg[  -\sqrt{s_{th} -s} \operatorname{Res}g(s_{th}), \nonumber \\ 
          &{} - \frac{(s_{th} -s)^{3/2}}{\pi}\int_{s_{th}}^{\infty}  \frac{\ln\frac{|F_Q^{\pi}(z)|}{F_Q^{\pi}(s_{th})}  \ dz}{(z-s_{th})^{3/2}(z-s)}   \Bigg], \\
   \delta(s) &{}= \phi(s) +\sqrt{s -s_{th}} \operatorname{Res}g(s_{th}) \nonumber \\ &{} 
              -   \frac{(s -s_{th})^{3/2}}{\pi}\operatorname{PV}\int_{s_{th}}^{\infty}  \frac{\ln\frac{|F_Q^{\pi}(z)|}{F_Q^{\pi}(s_{th})}  \ dz}{(z-s_{th})^{3/2}(z-s)}.  \label{eq:phaseZeroDRS} 
\end{align}
Similarly, the analogous of DR2 reads
\begin{align}
   F_{Q}^{\pi}(s) &{}= F_{Q}^{\pi}(s_{th})  B(s) \left[ F_{Q}^{\pi}(s_{th}) B(0) \right]^{-\left(\frac{ s_{th} -s}{s_{th} }\right)^{3/2}} \times \nonumber\\ 
          &{}   \operatorname{exp}\Bigg[  -\frac{s}{s_{th} }\sqrt{s_{th} -s} \operatorname{Res}g(s_{th}) \nonumber\\ 
          &{} - \frac{s(s_{th} -s)^{3/2}}{\pi}\int_{s_{th}}^{\infty}  \frac{\ln\frac{|F_Q^{\pi}(z)|}{F_Q^{\pi}(s_{th})}  \ dz}{z(z-s_{th})^{3/2}(z-s)}   \Bigg], \label{eq:DR2Zero} \\
   \delta(s) &{}= \phi(s) +\frac{s}{s_{th}}\sqrt{s -s_{th}} \operatorname{Res}g(s_{th}) \nonumber \\ 
          &{}  -\left( \frac{s -s_{th}}{s_{th}} \right)^{3/2} \ln[F_{Q}^{\pi}(s_{th}) B(0)] \nonumber \\ 
          &{}   -\frac{(s -s_{th})^{3/2}}{\pi}\operatorname{PV}\int_{s_{th}}^{\infty}  \frac{\ln\frac{|F_Q^{\pi}(z)|}{F_Q^{\pi}(s_{th})}  \ dz}{(z-s_{th})^{3/2}z(z-s)}. \label{eq:phaseZeroDR2}
\end{align}
Finally, the sum rule in \cref{eq:TheSumRule} is modified as 
\begin{equation}\label{eq:TheSumRule0}
    \frac{\sqrt{s_{th}} \operatorname{Res}g(s_{th})+ 
    \frac{s_{th}^{3/2}}{\pi} \int_{s_{th}}^{\infty} \frac{\ln\frac{|F_Q^{\pi}(z)|}{F_Q^{\pi}(s_{th})}  \ dz}{z(z -s_{th})^{3/2}} }{\ln[F_{Q}^{\pi}(s_{th})  B(0)  ]} =1.
\end{equation}
Note that, once more, such a sum rule is also relevant for the asymptotics of \cref{eq:DR2Zero}.
Finally, the slope is modified as 
\begin{multline}
  b_{\pi} = \frac{B'(0)}{B(0)} -\frac{\operatorname{Res}g(s_{th})}{\sqrt{s_{th}}}  + \frac{3}{2}\frac{\ln[ F_Q^{\pi}(s_{th}) B(0) ]}{s_{th}}
    \\ -\frac{s_{th}^{3/2}}{\pi}\int_{s_{th}}^{\infty} \frac{\ln|F_Q^{\pi}(z)/F_Q^{\pi}(s_{th})|}{z^2(z-s_{th})^{3/2}}. \label{eq:slopeZEROdrs}
\end{multline}
Once again, the modifications implied by the presence of zeros alter the value/phase of the form factor below/above threshold. 
In addition, it can be shown that the terms shifting the phase obtained from DR2 lead precisely to the additional terms in the 
twice-subtracted DR in the presence of zeros, see \cref{eq:DRphase2S0}.

\begin{figure}[h]
    \centering
\includegraphics[width=0.48\textwidth]{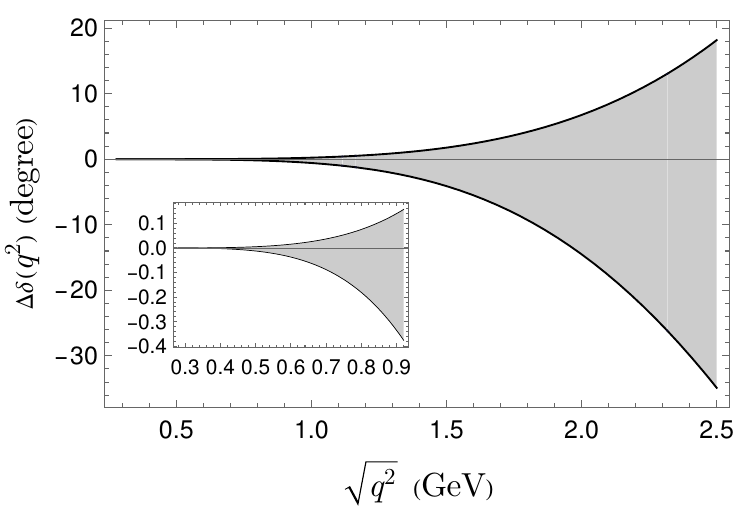} 
    \caption{The systematic uncertainty from potential zeros belonging to the class (i) boundary.}
    \label{fig:PhaseZero}
\end{figure}

As a consequence, we have two options to test for the presence of zeros: (i) to look at the variation in the spacelike region 
(ii) to check the validity of the sum rule in \cref{eq:TheSumRule0}. It turns out that, for most regions in the complex plane, 
it is the second option that leads to stronger bounds. In particular, from our results and considerations in the main text, 
such a sum rule can be stated as 
\begin{multline}
    \frac{ 
    s_{th}^{3/2} \int_{s_{th}}^{\Lambda} \frac{\ln\frac{|F_Q^{\pi}(z)|}{F_Q^{\pi}(s_{th})}  \ dz}{z(z -s_{th})^{3/2}} }{\pi \ln F_{Q}^{\pi}(s_{th})} > \\
    \frac{  \ln[F_{Q}^{\pi}(s_{th}) B(0) ] -\sqrt{s_{th}} \operatorname{Res}g(s_{th})  }{ \ln F_{Q}^{\pi}(s_{th})  }.\label{eq:sumruleineq}
\end{multline}  
for sufficiently large $\Lambda$. In particular, we can take our value $1.003$ for $\Lambda=3$~GeV for the LHS
(allowing for extrapolation to higher $\Lambda$ would provide more stringent bounds).
In addition, one can use the model-independent bounds derived in Ref.~\cite{Ananthanarayan:2011xt}.
\begin{figure}[h]
    \centering
\includegraphics[width=0.48\textwidth]{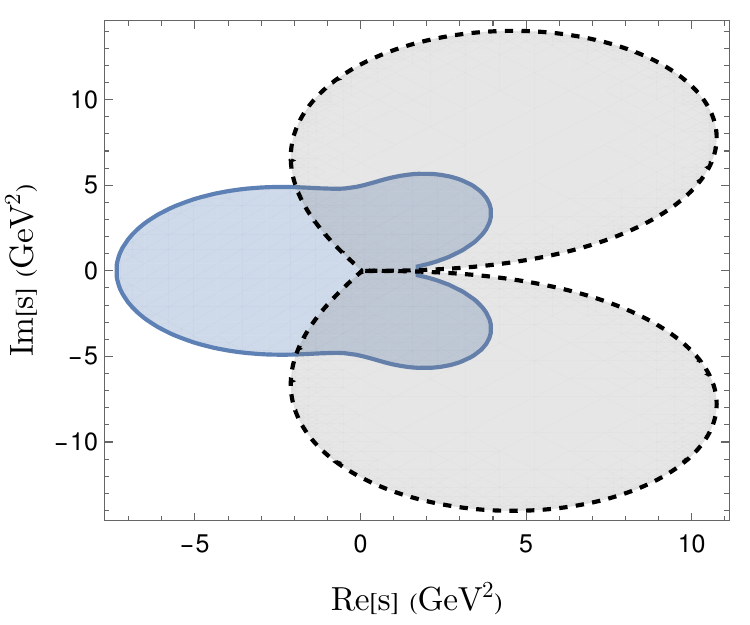} 
    \caption{The (shaded) regions where zeros are excluded. The gray-dotted region follows from \cref{eq:sumruleineq}, 
    whereas the blue-thick region follows from Ref.~\cite{Ananthanarayan:2011xt} (cf. their Fig.~4).}
    \label{fig:ZeroExcl}
\end{figure}
Both are shown in \cref{fig:ZeroExcl}. Note in particular that Ref.~\cite{Ananthanarayan:2011xt} excludes zeros near the timelike 
axis below $0.846~\textrm{GeV}^2$ approximately. 
In the following, we discuss the corrections that such hypothetical zeros outside the excluded regions have on the phase. 
In particular, we take the shift required in \cref{eq:phaseZeroDR2}, since it is the DR employed in this work and because it leads 
to the milder corrections. To do so, we must we distinguish two class of boundaries in \cref{fig:ZeroExcl}: 
(i) those in the spacelike region and the upper timelike boundary and 
(ii) those in the lower timelike boundary. 
Regarding class (i), those configurations leading to larger uncertainties come from the bounds in \cite{Ananthanarayan:2011xt} 
at $z=(-7.3+0.86i)~\textrm{GeV}^2$ and $z=(-1.95+ 4.89i)~\textrm{GeV}^2$ (complex conjugates poles are implied)  and are shown in \cref{fig:PhaseZero}. 
Regarding class (ii), such zeros generally feature a steep (steeper as they approach the real axis) rise of about $2\pi$ in the phase, 
which is easily understood from the argument theorem.   
We don't find a strong physical motivation for an arbitrary number of such shifts, completely distorting our result. 
At most, we would expect them in the neighborhood of some resonance 
(see for instance the original phase in our interpolating model, that seems to approach $3\pi$; 
note however that the original model features two poles which are not complex-conjugate, as it 
does not fulfill the Schwarz reflection principle, and $|F_Q^{\pi}(s +i\epsilon)| \neq |F_Q^{\pi}(s -i\epsilon)|$. 
Further, the argument theorem should be modified, as the discontinuity differs from $2\delta(\Lambda^2)$).

\begin{figure*}
    \centering
\includegraphics[width=0.48\textwidth]{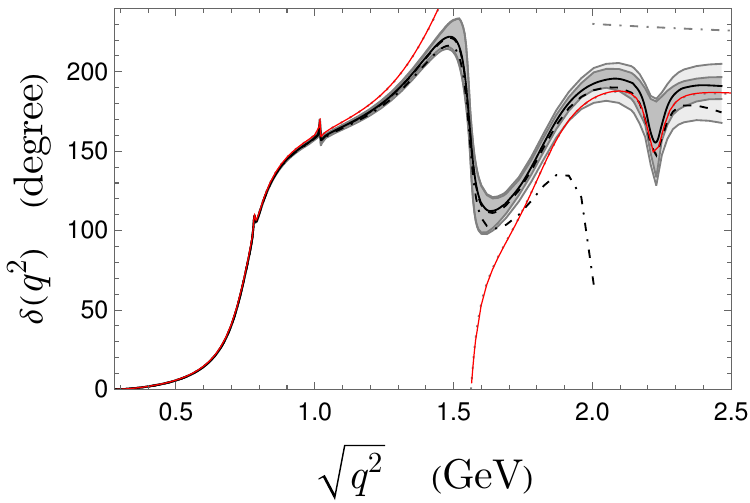} 
\includegraphics[width=0.48\textwidth]{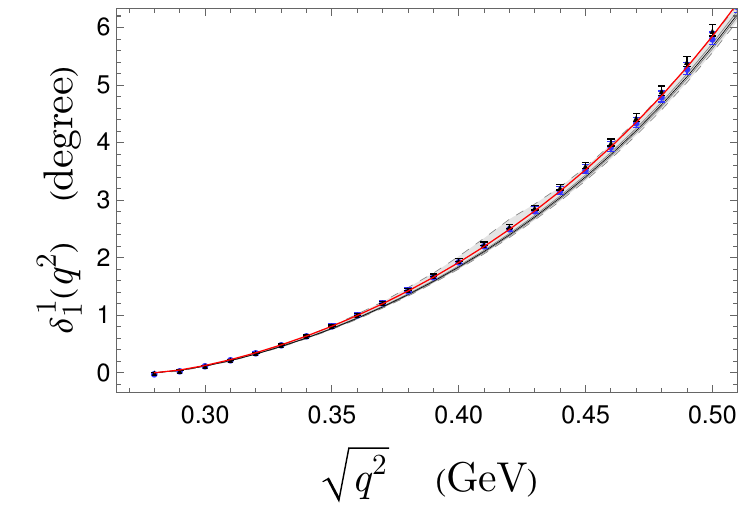} 
\includegraphics[width=0.48\textwidth]{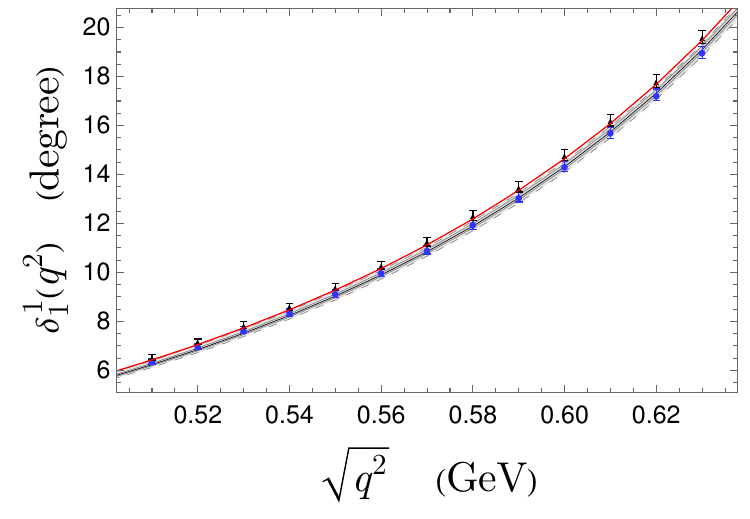} 
\includegraphics[width=0.48\textwidth]{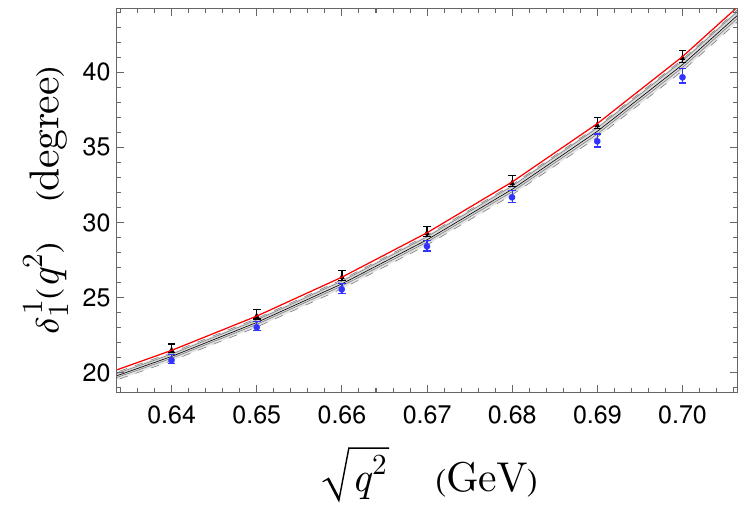} 
\includegraphics[width=0.48\textwidth]{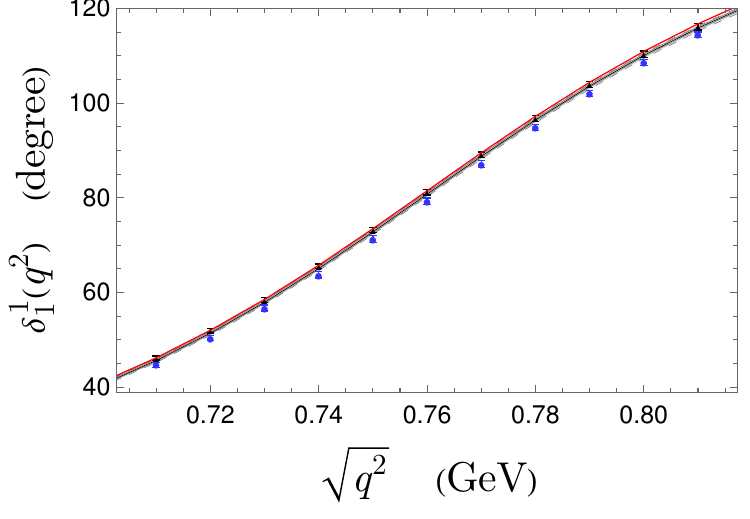} 
\includegraphics[width=0.48\textwidth]{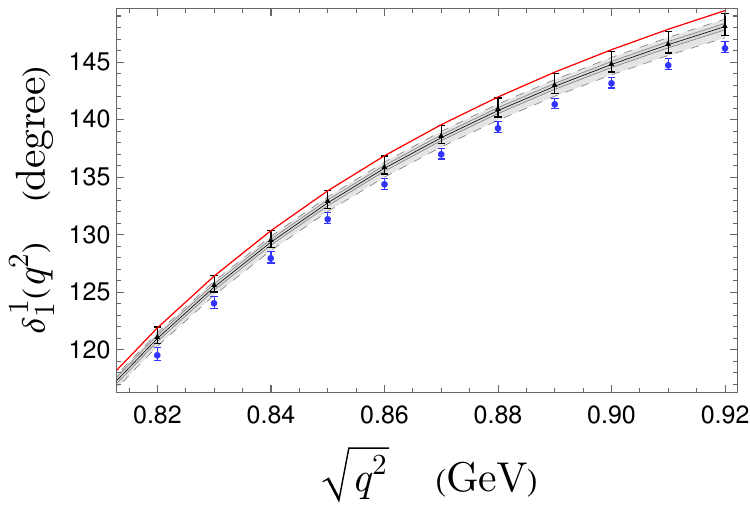} 
    \caption{We compare the result of a unitarization model based on Cauchy's theorem for $\operatorname{Im} F_Q^{\pi}(s)$ that
    features nearby complex-conjugated zeroes (red line), as described in the main text, to 
    the results of the DR2 for the full phase (upper-left panel; cf. \cref{fig:PhaseReIm} and the corresponding coloring scheme) 
    and isovector phase (other panels; cf. \cref{fig:PhaseShift}  and the corresponding coloring scheme).\label{fig:PhaseShiftCauchy}} 
\end{figure*}

To further explore this scenario and show its impact, 
we explore what happens if the absence of zeroes is not imposed 
in the formalism. One possibility is to unitarize our GS model by using 
Cauchy's integral representation for $\operatorname{Im}F_{Q}^{\pi}(z)$. 
To do so, we remove the imaginary part at threshold and employ Cauchy's 
integral theorem to find the real part. In doing so, we find 
the phase in the upper-left panel in \cref{fig:PhaseShiftCauchy}, that 
essentially reproduces the original one shown in \cref{fig:PhaseReIm} 
and features complex-conjugate poles at $s=(1.56\pm 0.25i)^2~\textrm{GeV}^2$, 
offering an example for the quoted $2\pi$ jumps. 
The corresponding phase for the isovector form factor is shown in \cref{fig:PhaseShiftCauchy} 
and illustrates the potential systematics induced by 
nearby zeros with some potential physical motivation (rather than arbitrary zeros).
Of course, this does not imply the existence of such a zero, as it is just 
a model (for instance, the GS model lacks important two-body contributions 
to the imaginary part, such as $K\bar{K}$, $\omega\pi$, or $a_1\pi$ ones, 
that necessarily change the phase). 
Actually, the phase for the resonant model in 
Ref.~\cite{Gonzalez-Solis:2019iod} lacks such a jump, which illustrates 
the ambiguities of inverse problems and potential large model-dependencies. 
See also the study in Ref.~\cite{Moussallam:2007qc} 
for related discussions in the context of $K\pi$ form factors, where solutions 
asymptotically approaching either $\pi$ or $3\pi$ are obtained. 
This is in marked contrast with the case of the dispersion 
relations here employed that, for the potential scenario in which the form factor 
is actually devoid of zeros, show a large stability with respect to mid- and high-energy 
input as illustrated in \cref{app:syst} (see \cref{fig:DiscretePhase}), 
producing then robust results.

To advance in this problem, it would be interesting to better understand 
the region beyond 1~GeV, including for instance the $\omega\pi$ or $K^+K^-$ 
electromagnetic form factors.

\section{Systematic errors from data interpolation\label{app:syst}}

As mentioned in \cref{sec:defs}, DR2 shows a potentially large 
sensitivity to the form factor behavior close to threshold, see \cref{eq:SubtThZero,eq:SubtThZeroPhase}. 
Given that, to some extent, we rely on the extrapolation of the chosen model, 
one may wonder about model dependencies and potential systematic uncertainties. 
To estimate them, we will investigate variations of the form factor behavior close to threshold. 

Clearly, the largest impact comes from variations of $\ln F_Q^{\pi}(s_{th})$ 
[cf. the prefactor in \cref{eq:SubtThZero} and the first term 
in \cref{eq:SubtThZeroPhase}], that can be easily estimated varying 
$\ln F_Q^{\pi}(s_{th})$ while keeping $F_Q^{\pi}(s)/F_Q^{\pi}(s_{th})$ 
fixed. Further, in order to investigate energy-dependent variations, we will introduce 
the following threshold expansion
\begin{equation}\label{eq:ffthmodel}
  \frac{F_Q^{\pi}(s)}{F_Q^{\pi}(s_{th})} = 1 + \frac{\alpha_1 s}{1 -\alpha_2 s} 
   +  \frac{ i\beta \frac{s^{3/2}}{M_{\rho}^2\sqrt{s_{th}}}  }{1 -\frac{s}{M_{\rho}^2}} , 
\end{equation}
which is able to describe our model from threshold up to around
$0.42$~GeV ---arguably enough to assess potential large dependencies 
to the form factor close to threshold.
However, $\ln F_Q^{\pi}(s_{th})$ or the parameters in \cref{eq:ffthmodel} 
cannot be chosen at will, for they (in general) imply a violation of the sum rule in \cref{eq:TheSumRule}.
Indeed, as we will see, in the absence of zeros the sum rule greatly reduces the sensitivity 
to the form factor behavior close to threshold. 

For instance, if varying $\ln F_Q^{\pi}(s_{th})$ alone, we find that only 
the $ F_Q^{\pi}(s_{th})\in (1.1737,1.1747)$ range is allowed when
accounting for the limits discussed in \cref{sec:testSR,sec:realdata}. 
This immediately translates into a relative uncertainty 
$\left( F_Q^{\pi}(s_{th}) / F_Q^{\pi}(s_{th}) \vert_{\textrm{fit}} \right)^{1 - (1 -s/s_{th})^{3/2}}$
for the form factor below threshold and a phase variation of 
$-\ln\left(F_Q^{\pi}(s_{th}) / F_Q^{\pi}(s_{th}) \vert_{\textrm{fit}} \right) \left(s/s_{th} -1\right)^{3/2}$ 
above.  
At this stage, two remarks are relevant. First, it 
must be emphasized that, nonetheless, part of this error appears already 
in our statistical error budget since our MC sample contains limiting 
cases for the sum rule. Second, potential zeros discussed in 
\cref{app:zeros} could slightly shift these bounds, that rigorously holds in the absence of zeros.  

Further, to vary the parameters in \cref{eq:ffthmodel}, we first fix the 
value for $F_Q^{\pi}(s_{th})$; then we vary the parameters in regions allowed 
by the sum rule. We note that there is a hierarchy when varying them. 
In order of relevance, we find $\{ F_Q^{\pi}(s_{th}), \alpha_1, \alpha_2, \beta \}$ ,
with $\beta$ playing a marginal role.  
For $ F_Q^{\pi}(s_{th})\in (1.1737,1.1747)$, their variation leads 
to additional systematics beyond the bounds that could be obtained 
if varying $F_Q^{\pi}(s_{th})$ alone. Further, since we have the 
independent estimate $F_Q^{\pi}(s_{th}) =1.176(2)$~\cite{Ananthanarayan:2018nyx}, 
we also consider this case that, for instance, leads to the larger 
systematics for the phase close tho threshold. 
This, together with the systematics from variations of 
$ F_Q^{\pi}(s_{th})\in (1.1737,1.1747)$  
will be considered as the systematic error uncertainty. 

\begin{figure}[h]
    \centering
\includegraphics[width=0.48\textwidth]{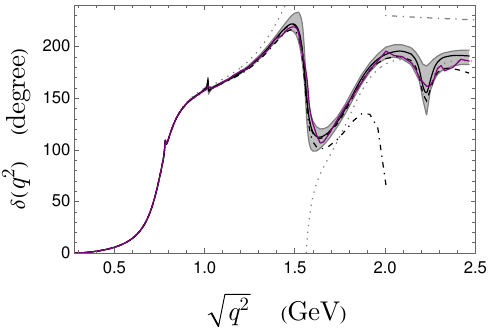} 
    \caption{The result from DR2 if switching to a linear 
    interpolation of BaBar's data above $0.6$~GeV (purple) 
    compared to the results in \cref{fig:PhaseReIm} (same color scheme).}
    \label{fig:DiscretePhase}
\end{figure}

Finally, to close on systematic uncertainties, we display in 
\cref{fig:DiscretePhase} the result that would be obtained for the phase if 
replacing our fitting function above $\sqrt{s}=0.6$~GeV by 
a simple linear interpolation to BaBar's data. 
Indeed, it would be possible to choose 
a cutoff as low as $.45$~GeV, leading to very similar 
results for the phase above $0.6$~GeV. This emphasizes 
the stability of this approach with respect to the particular 
parameterization chosen from energies as low as $0.6$~GeV.

\section{Numerical results\label{app:datasets}}

In this appendix, we provide the form factor (modulus and phase) for different points above the $\pi\pi$ threshold. We include in addition the shown bands at 68\% CL. We note however that the upper and lower bands do not necessarily identify with particular values of the fitting parameters ---in order to perform a careful error analysis, accounting for correlations, any quantity derived from our results should be obtained from the set consisting of all pseudofits. In addition, we provide (in parenthesis) the systematic uncertainty for the phase
corresponding to the interpolation uncertainties discussed in \cref{app:syst}. The systematic uncertainties corresponding to potential zeros must be considered separately, as it largely depends on their position. See our discussion in \cref{app:zeros}.

\begin{table}
\caption[t]{The value for the modulus and phase of the 
electromagnetic form factor of the charged pion $F_Q^{\pi}(s) = |F_Q^{\pi}(s)| e^{i\delta_Q(s)}$. 
The error for the modulus  stems from the fit and is statistical only. 
The error for the phase is split into the statistical and systematic (in parenthesis) one. 
The latter accounting from the interpolation uncertainties exclusively, 
whereas the systematic of zeroes require, in general, a separate treatment 
(cf. \cref{app:zeros}).   
}
\begin{tabular}{ccc}
\toprule
$\sqrt{s}$~(GeV)  &  $|F_Q^{\pi}(s)|$  &  $\delta_Q(s) (^{\circ})$ \\\midrule
$  0.279$ & $1.174_{-0.0009}^{0.001}$ & $0$ \\
$  0.299$ & $1.209_{-0.001}^{0.001}$ & $0.11_{-0.001(0.002)}^{0.001(0.001)}$ \\
$  0.319$ & $1.246_{-0.001}^{0.002}$ & $0.316_{-0.003(0.005)}^{0.004(0.009)}$ \\
$  0.339$ & $1.288_{-0.001}^{0.002}$ & $0.595_{-0.005(0.01)}^{0.007(0.03)}$ \\
$  0.359$ & $1.334_{-0.002}^{0.002}$ & $0.93_{-0.008(0.02)}^{0.01(0.07)}$ \\
$  0.379$ & $1.386_{-0.002}^{0.003}$ & $1.34_{-0.01(0.02)}^{0.01(0.1)}$ \\
$  0.399$ & $1.443_{-0.002}^{0.003}$ & $1.82_{-0.02(0.03)}^{0.02(0.2)}$ \\
$  0.419$ & $1.509_{-0.002}^{0.003}$ & $2.38_{-0.02(0.04)}^{0.02(0.3)}$ \\
$  0.439$ & $1.58_{-0.003}^{0.004}$ & $3.03_{-0.02(0.05)}^{0.03(0.2)}$ \\
$  0.459$ & $1.664_{-0.003}^{0.004}$ & $3.76_{-0.03(0.07)}^{0.04(0.2)}$ \\
$  0.479$ & $1.76_{-0.003}^{0.005}$ & $4.64_{-0.04(0.08)}^{0.04(0.2)}$ \\
$  0.499$ & $1.866_{-0.004}^{0.006}$ & $5.6_{-0.04(0.1)}^{0.05(0.2)}$ \\
$  0.519$ & $1.992_{-0.004}^{0.006}$ & $6.78_{-0.05(0.1)}^{0.06(0.2)}$ \\
$  0.52$ & $1.998_{-0.004}^{0.006}$ & $6.84_{-0.05(0.1)}^{0.06(0.2)}$ \\
$  0.535$ & $2.1_{-0.005}^{0.007}$ & $7.91_{-0.06(0.1)}^{0.07(0.2)}$ \\
$  0.55$ & $2.224_{-0.005}^{0.008}$ & $9.04_{-0.07(0.1)}^{0.08(0.2)}$ \\
$  0.565$ & $2.358_{-0.006}^{0.009}$ & $10.35_{-0.07(0.2)}^{0.09(0.2)}$ \\
$  0.58$ & $2.52_{-0.007}^{0.01}$ & $11.9_{-0.08(0.2)}^{0.1(0.2)}$ \\
$  0.595$ & $2.69_{-0.007}^{0.01}$ & $13.7_{-0.09(0.2)}^{0.1(0.2)}$ \\
$  0.61$ & $2.9_{-0.009}^{0.01}$ & $15.7_{-0.1(0.2)}^{0.1(0.2)}$ \\
$  0.625$ & $3.14_{-0.01}^{0.01}$ & $18.2_{-0.1(0.2)}^{0.1(0.2)}$ \\
$  0.64$ & $3.42_{-0.01}^{0.02}$ & $21.1_{-0.1(0.3)}^{0.1(0.2)}$ \\
$  0.646$ & $3.54_{-0.01}^{0.02}$ & $22.4_{-0.1(0.3)}^{0.2(0.2)}$ \\
$  0.652$ & $3.66_{-0.01}^{0.02}$ & $23.8_{-0.1(0.3)}^{0.2(0.2)}$ \\
$  0.658$ & $3.8_{-0.01}^{0.02}$ & $25.4_{-0.2(0.3)}^{0.2(0.2)}$ \\
$  0.664$ & $3.96_{-0.01}^{0.02}$ & $27._{-0.2(0.3)}^{0.2(0.2)}$ \\
$  0.67$ & $4.1_{-0.01}^{0.02}$ & $28.8_{-0.2(0.3)}^{0.2(0.2)}$ \\
$  0.676$ & $4.28_{-0.01}^{0.02}$ & $30.8_{-0.2(0.3)}^{0.2(0.2)}$ \\
\end{tabular}
\end{table}


\begin{tabular}{ccc}
$  0.682$ & $4.44_{-0.01}^{0.02}$ & $33._{-0.2(0.3)}^{0.2(0.2)}$ \\
$  0.688$ & $4.64_{-0.02}^{0.02}$ & $35.4_{-0.2(0.3)}^{0.2(0.2)}$ \\
$  0.694$ & $4.82_{-0.02}^{0.02}$ & $37.8_{-0.2(0.4)}^{0.2(0.2)}$ \\
$  0.7$ & $5.02_{-0.02}^{0.02}$ & $40.6_{-0.2(0.4)}^{0.2(0.3)}$ \\
$  0.706$ & $5.22_{-0.02}^{0.03}$ & $43.5_{-0.2(0.4)}^{0.3(0.3)}$ \\
$  0.71$ & $5.37_{-0.02}^{0.03}$ & $45.6_{-0.2(0.4)}^{0.3(0.3)}$ \\
$  0.714$ & $5.49_{-0.02}^{0.03}$ & $48._{-0.2(0.4)}^{0.3(0.3)}$ \\
$  0.718$ & $5.64_{-0.02}^{0.03}$ & $50.1_{-0.2(0.4)}^{0.3(0.3)}$ \\
$  0.722$ & $5.76_{-0.02}^{0.03}$ & $52.8_{-0.2(0.4)}^{0.3(0.3)}$ \\
$  0.726$ & $5.88_{-0.02}^{0.03}$ & $55.2_{-0.2(0.4)}^{0.3(0.3)}$ \\
$  0.73$ & $6.03_{-0.02}^{0.03}$ & $57.9_{-0.2(0.4)}^{0.3(0.3)}$ \\
$  0.734$ & $6.12_{-0.02}^{0.03}$ & $60.9_{-0.2(0.4)}^{0.3(0.3)}$ \\
$  0.738$ & $6.24_{-0.02}^{0.03}$ & $63.6_{-0.2(0.4)}^{0.3(0.3)}$ \\
$  0.742$ & $6.33_{-0.02}^{0.03}$ & $66.6_{-0.3(0.4)}^{0.3(0.3)}$ \\
$  0.746$ & $6.42_{-0.02}^{0.03}$ & $69.9_{-0.3(0.5)}^{0.3(0.3)}$ \\
$  0.75$ & $6.48_{-0.02}^{0.03}$ & $72.9_{-0.2(0.5)}^{0.3(0.3)}$ \\
$  0.754$ & $6.54_{-0.02}^{0.03}$ & $76.2_{-0.2(0.5)}^{0.3(0.3)}$ \\
$  0.758$ & $6.57_{-0.02}^{0.03}$ & $79.5_{-0.2(0.5)}^{0.3(0.3)}$ \\
$  0.76$ & $6.57_{-0.02}^{0.03}$ & $81.3_{-0.2(0.5)}^{0.3(0.3)}$ \\
$  0.764$ & $6.6_{-0.02}^{0.03}$ & $84.3_{-0.2(0.5)}^{0.3(0.3)}$ \\
$  0.767$ & $6.63_{-0.02}^{0.03}$ & $87.3_{-0.3(0.5)}^{0.3(0.3)}$ \\
$  0.771$ & $6.66_{-0.02}^{0.03}$ & $90.6_{-0.3(0.5)}^{0.3(0.3)}$ \\
$  0.774$ & $6.72_{-0.02}^{0.04}$ & $94.5_{-0.3(0.5)}^{0.3(0.3)}$ \\
$  0.778$ & $6.72_{-0.03}^{0.04}$ & $100.2_{-0.3(0.5)}^{0.3(0.3)}$ \\
$  0.779$ & $6.68_{-0.03}^{0.04}$ & $102._{-0.3(0.5)}^{0.4(0.3)}$ \\
$  0.78$ & $6.6_{-0.02}^{0.04}$ & $104._{-0.3(0.5)}^{0.4(0.3)}$ \\
$  0.781$ & $6.48_{-0.03}^{0.04}$ & $106._{-0.3(0.5)}^{0.4(0.3)}$ \\
$  0.782$ & $6.24_{-0.03}^{0.04}$ & $108._{-0.3(0.5)}^{0.4(0.3)}$ \\
$  0.782$ & $6._{-0.03}^{0.04}$ & $108.8_{-0.3(0.5)}^{0.4(0.3)}$ \\
$  0.783$ & $5.76_{-0.03}^{0.04}$ & $109.2_{-0.2(0.5)}^{0.4(0.3)}$ \\
$  0.784$ & $5.56_{-0.03}^{0.04}$ & $108.4_{-0.2(0.5)}^{0.4(0.3)}$ \\
$  0.785$ & $5.48_{-0.03}^{0.04}$ & $107.6_{-0.2(0.5)}^{0.4(0.3)}$ \\
$  0.788$ & $5.37_{-0.03}^{0.03}$ & $106._{-0.4(0.5)}^{0.4(0.4)}$ \\
$  0.791$ & $5.38_{-0.02}^{0.02}$ & $106._{-0.3(0.6)}^{0.4(0.4)}$ \\
$  0.794$ & $5.38_{-0.02}^{0.02}$ & $107.2_{-0.3(0.6)}^{0.4(0.4)}$ \\
$  0.797$ & $5.34_{-0.02}^{0.02}$ & $108.8_{-0.3(0.6)}^{0.4(0.4)}$ \\
\end{tabular}


\begin{tabular}{ccc}
$  0.8$ & $5.28_{-0.02}^{0.02}$ & $110.8_{-0.3(0.6)}^{0.4(0.4)}$ \\
$  0.803$ & $5.2_{-0.02}^{0.02}$ & $112.2_{-0.3(0.6)}^{0.3(0.4)}$ \\
$  0.806$ & $5.12_{-0.02}^{0.02}$ & $114._{-0.3(0.6)}^{0.3(0.4)}$ \\
$  0.809$ & $5.04_{-0.02}^{0.02}$ & $115.5_{-0.3(0.6)}^{0.3(0.4)}$ \\
$  0.812$ & $4.96_{-0.02}^{0.02}$ & $117._{-0.3(0.6)}^{0.3(0.4)}$ \\
$  0.815$ & $4.86_{-0.02}^{0.02}$ & $118.8_{-0.3(0.6)}^{0.3(0.4)}$ \\
$  0.818$ & $4.78_{-0.02}^{0.02}$ & $120._{-0.3(0.6)}^{0.3(0.4)}$ \\
$  0.82$ & $4.72_{-0.02}^{0.02}$ & $121.2_{-0.3(0.6)}^{0.3(0.4)}$ \\
$  0.827$ & $4.52_{-0.02}^{0.02}$ & $124.2_{-0.3(0.6)}^{0.3(0.4)}$ \\
$  0.833$ & $4.32_{-0.02}^{0.02}$ & $126.9_{-0.2(0.7)}^{0.3(0.4)}$ \\
$  0.84$ & $4.12_{-0.02}^{0.02}$ & $129.3_{-0.2(0.7)}^{0.3(0.4)}$ \\
$  0.846$ & $3.94_{-0.02}^{0.02}$ & $131.7_{-0.2(0.7)}^{0.3(0.4)}$ \\
$  0.853$ & $3.78_{-0.01}^{0.02}$ & $133.8_{-0.2(0.7)}^{0.3(0.5)}$ \\
$  0.86$ & $3.6_{-0.01}^{0.02}$ & $135.6_{-0.2(0.7)}^{0.3(0.5)}$ \\
$  0.866$ & $3.46_{-0.01}^{0.02}$ & $137.4_{-0.2(0.8)}^{0.3(0.5)}$ \\
$  0.87$ & $3.38_{-0.01}^{0.02}$ & $138.6_{-0.2(0.8)}^{0.3(0.5)}$ \\
$  0.88$ & $3.16_{-0.01}^{0.02}$ & $140.7_{-0.2(0.8)}^{0.3(0.5)}$ \\
$  0.89$ & $2.98_{-0.01}^{0.01}$ & $143.1_{-0.2(0.8)}^{0.3(0.5)}$ \\
$  0.9$ & $2.81_{-0.01}^{0.01}$ & $144.9_{-0.2(0.9)}^{0.3(0.5)}$ \\
$  0.91$ & $2.66_{-0.01}^{0.01}$ & $146.7_{-0.3(0.9)}^{0.3(0.6)}$ \\
$  0.92$ & $2.52_{-0.01}^{0.01}$ & $148.2_{-0.3(0.9)}^{0.3(0.6)}$ \\
$  0.93$ & $2.4_{-0.009}^{0.01}$ & $149.7_{-0.3(1.)}^{0.3(0.6)}$ \\
$  0.94$ & $2.28_{-0.009}^{0.01}$ & $150.9_{-0.3(1.)}^{0.3(0.6)}$ \\
$  0.95$ & $2.18_{-0.008}^{0.01}$ & $152.4_{-0.3(1.)}^{0.4(0.7)}$ \\
$  0.96$ & $2.08_{-0.008}^{0.01}$ & $153.6_{-0.3(1.)}^{0.4(0.7)}$ \\
$  0.97$ & $2._{-0.008}^{0.01}$ & $154.8_{-0.4(1.)}^{0.4(0.7)}$ \\
$  0.98$ & $1.92_{-0.008}^{0.01}$ & $155.5_{-0.4(1.)}^{0.5(0.7)}$ \\
$  1.01$ & $1.73_{-0.01}^{0.01}$ & $160._{-1.(1.)}^{1.(0.8)}$ \\
$  1.014$ & $1.7_{-0.03}^{0.02}$ & $162._{-2.(1.)}^{2.(0.8)}$ \\
$  1.015$ & $1.71_{-0.03}^{0.03}$ & $162._{-2.(1.)}^{2.(0.8)}$ \\
$  1.016$ & $1.68_{-0.04}^{0.04}$ & $164._{-2.(1.)}^{2.(0.8)}$ \\
$  1.017$ & $1.68_{-0.06}^{0.06}$ & $165._{-3.(1.)}^{3.(0.8)}$ \\
$  1.018$ & $1.62_{-0.09}^{0.09}$ & $168._{-3.(1.)}^{3.(0.8)}$ \\
$  1.019$ & $1.5_{-0.1}^{0.1}$ & $168._{-3.(1.)}^{2.(0.8)}$ \\
$  1.02$ & $1.4_{-0.1}^{0.1}$ & $164._{-3.(1.)}^{1.(0.8)}$ \\
$  1.021$ & $1.4_{-0.08}^{0.1}$ & $159._{-4.(1.)}^{3.(0.8)}$ \\
\end{tabular}


\begin{tabular}{ccc}
$  1.022$ & $1.47_{-0.05}^{0.07}$ & $156._{-3.(1.)}^{3.(0.8)}$ \\
$  1.023$ & $1.5_{-0.03}^{0.05}$ & $156._{-3.(1.)}^{3.(0.8)}$ \\
$  1.024$ & $1.5_{-0.02}^{0.03}$ & $158._{-2.(1.)}^{2.(0.8)}$ \\
$  1.025$ & $1.53_{-0.01}^{0.03}$ & $158._{-2.(1.)}^{2.(0.8)}$ \\
$  1.04$ & $1.51_{-0.006}^{0.01}$ & $160._{-0.7(1.)}^{0.8(0.9)}$ \\
$  1.07$ & $1.39_{-0.006}^{0.01}$ & $162.6_{-0.6(2.)}^{0.6(1.)}$ \\
$  1.1$ & $1.27_{-0.006}^{0.01}$ & $165.2_{-0.6(2.)}^{0.7(1.)}$ \\
$  1.13$ & $1.179_{-0.007}^{0.009}$ & $168._{-0.7(2.)}^{0.8(1.)}$ \\
$  1.16$ & $1.098_{-0.007}^{0.009}$ & $171._{-0.8(2.)}^{0.9(1.)}$ \\
$  1.19$ & $1.017_{-0.007}^{0.009}$ & $174._{-0.9(2.)}^{1.(1.)}$ \\
$  1.2$ & $0.999_{-0.007}^{0.009}$ & $175._{-1.(2.)}^{1.(1.)}$ \\
$  1.238$ & $0.909_{-0.008}^{0.009}$ & $180._{-1.(2.)}^{1.(2.)}$ \\
$  1.276$ & $0.828_{-0.009}^{0.009}$ & $186._{-1.(3.)}^{2.(2.)}$ \\
$  1.314$ & $0.747_{-0.009}^{0.009}$ & $192._{-2.(3.)}^{2.(2.)}$ \\
$  1.352$ & $0.65_{-0.01}^{0.01}$ & $198._{-2.(3.)}^{2.(2.)}$ \\
$  1.39$ & $0.54_{-0.01}^{0.01}$ & $207._{-3.(4.)}^{3.(2.)}$ \\
$  1.428$ & $0.41_{-0.01}^{0.01}$ & $216._{-4.(4.)}^{3.(2.)}$ \\
$  1.43$ & $0.41_{-0.01}^{0.01}$ & $216._{-4.(4.)}^{4.(2.)}$ \\
$  1.435$ & $0.39_{-0.01}^{0.01}$ & $216._{-4.(4.)}^{4.(2.)}$ \\
$  1.44$ & $0.37_{-0.01}^{0.01}$ & $216._{-4.(4.)}^{4.(2.)}$ \\
$  1.445$ & $0.36_{-0.01}^{0.01}$ & $216._{-4.(4.)}^{4.(2.)}$ \\
$  1.45$ & $0.34_{-0.01}^{0.01}$ & $220._{-4.(4.)}^{5.(3.)}$ \\
$  1.455$ & $0.32_{-0.01}^{0.01}$ & $220._{-5.(4.)}^{5.(3.)}$ \\
$  1.46$ & $0.31_{-0.01}^{0.01}$ & $220._{-5.(4.)}^{5.(3.)}$ \\
$  1.465$ & $0.3_{-0.01}^{0.02}$ & $222._{-5.(4.)}^{6.(3.)}$ \\
$  1.466$ & $0.28_{-0.01}^{0.02}$ & $222._{-5.(4.)}^{6.(3.)}$ \\
$  1.47$ & $0.28_{-0.009}^{0.02}$ & $222._{-5.(4.)}^{6.(3.)}$ \\
$  1.475$ & $0.26_{-0.009}^{0.02}$ & $224._{-6.(4.)}^{7.(3.)}$ \\
$  1.48$ & $0.24_{-0.01}^{0.02}$ & $224._{-6.(4.)}^{8.(3.)}$ \\
$  1.485$ & $0.22_{-0.01}^{0.02}$ & $224._{-7.(4.)}^{8.(3.)}$ \\
$  1.49$ & $0.22_{-0.01}^{0.02}$ & $225._{-7.(5.)}^{9.(3.)}$ \\
$  1.495$ & $0.2_{-0.009}^{0.02}$ & $220._{-7.(5.)}^{10.(3.)}$ \\
$  1.5$ & $0.18_{-0.01}^{0.02}$ & $220._{-8.(5.)}^{10.(3.)}$ \\
$  1.505$ & $0.16_{-0.01}^{0.02}$ & $220._{-9.(5.)}^{10.(3.)}$ \\
$  1.51$ & $0.16_{-0.01}^{0.02}$ & $220._{-9.(5.)}^{10.(3.)}$ \\
\end{tabular}


\begin{tabular}{ccc}
$  1.515$ & $0.14_{-0.01}^{0.02}$ & $220._{-10.(5.)}^{20.(3.)}$ \\
$  1.52$ & $0.12_{-0.01}^{0.02}$ & $220._{-10.(5.)}^{20.(3.)}$ \\
$  1.525$ & $0.12_{-0.01}^{0.02}$ & $220._{-10.(5.)}^{20.(3.)}$ \\
$  1.53$ & $0.1_{-0.02}^{0.02}$ & $200._{-10.(5.)}^{20.(3.)}$ \\
$  1.535$ & $0.1_{-0.02}^{0.02}$ & $200._{-10.(5.)}^{20.(3.)}$ \\
$  1.54$ & $0.09_{-0.02}^{0.03}$ & $210._{-10.(5.)}^{30.(3.)}$ \\
$  1.545$ & $0.06_{-0.02}^{0.03}$ & $180._{-10.(5.)}^{30.(3.)}$ \\
$  1.55$ & $0.06_{-0.03}^{0.03}$ & $180._{-10.(5.)}^{20.(3.)}$ \\
$  1.555$ & $0.06_{-0.03}^{0.03}$ & $180._{-10.(5.)}^{20.(3.)}$ \\
$  1.56$ & $0.06_{-0.03}^{0.03}$ & $160._{-10.(5.)}^{10.(3.)}$ \\
$  1.565$ & $0.06_{-0.03}^{0.03}$ & $150._{-20.(5.)}^{10.(3.)}$ \\
$  1.57$ & $0.06_{-0.03}^{0.03}$ & $140._{-20.(5.)}^{10.(3.)}$ \\
$  1.575$ & $0.09_{-0.03}^{0.03}$ & $140._{-20.(5.)}^{10.(3.)}$ \\
$  1.58$ & $0.09_{-0.02}^{0.03}$ & $130._{-20.(5.)}^{10.(3.)}$ \\
$  1.585$ & $0.09_{-0.02}^{0.03}$ & $130._{-20.(5.)}^{10.(3.)}$ \\
$  1.59$ & $0.1_{-0.02}^{0.02}$ & $120._{-20.(6.)}^{10.(3.)}$ \\
$  1.595$ & $0.12_{-0.02}^{0.02}$ & $120._{-20.(6.)}^{10.(3.)}$ \\
$  1.6$ & $0.12_{-0.02}^{0.02}$ & $120._{-20.(6.)}^{10.(3.)}$ \\
$  1.62$ & $0.16_{-0.01}^{0.02}$ & $117._{-10.(6.)}^{9.(4.)}$ \\
$  1.64$ & $0.2_{-0.01}^{0.02}$ & $112._{-10.(6.)}^{8.(4.)}$ \\
$  1.645$ & $0.22_{-0.01}^{0.02}$ & $112._{-10.(6.)}^{8.(4.)}$ \\
$  1.66$ & $0.24_{-0.009}^{0.02}$ & $112._{-9.(6.)}^{7.(4.)}$ \\
$  1.68$ & $0.28_{-0.008}^{0.02}$ & $114._{-7.(7.)}^{6.(4.)}$ \\
$  1.69$ & $0.3_{-0.009}^{0.02}$ & $120._{-7.(7.)}^{6.(4.)}$ \\
$  1.735$ & $0.39_{-0.01}^{0.01}$ & $125._{-5.(7.)}^{5.(4.)}$ \\
$  1.78$ & $0.45_{-0.02}^{0.01}$ & $140._{-5.(8.)}^{4.(5.)}$ \\
$  1.825$ & $0.47_{-0.02}^{0.01}$ & $156._{-5.(9.)}^{4.(5.)}$ \\
$  1.87$ & $0.44_{-0.01}^{0.02}$ & $168._{-5.(9.)}^{4.(6.)}$ \\
$  1.915$ & $0.4_{-0.009}^{0.02}$ & $180._{-5.(10.)}^{4.(6.)}$ \\
$  1.96$ & $0.34_{-0.008}^{0.02}$ & $190._{-4.(10.)}^{5.(6.)}$ \\
$  2.005$ & $0.28_{-0.01}^{0.01}$ & $196._{-4.(10.)}^{7.(7.)}$ \\
$  2.05$ & $0.234_{-0.02}^{0.009}$ & $196._{-5.(10.)}^{7.(7.)}$ \\
$  2.095$ & $0.2_{-0.02}^{0.01}$ & $198._{-8.(10.)}^{6.(8.)}$ \\
$  2.14$ & $0.16_{-0.02}^{0.02}$ & $192._{-10.(10.)}^{6.(8.)}$ \\
$  2.15$ & $0.16_{-0.02}^{0.02}$ & $192._{-10.(10.)}^{6.(8.)}$ \\
\end{tabular}


\begin{tabular}{ccc}
$  2.17$ & $0.12_{-0.02}^{0.03}$ & $189._{-10.(10.)}^{7.(9.)}$ \\
$  2.19$ & $0.1_{-0.02}^{0.05}$ & $180._{-10.(20.)}^{10.(9.)}$ \\
$  2.21$ & $0.12_{-0.03}^{0.04}$ & $160._{-10.(20.)}^{20.(9.)}$ \\
$  2.23$ & $0.18_{-0.03}^{0.03}$ & $150._{-20.(20.)}^{30.(9.)}$ \\
$  2.25$ & $0.24_{-0.06}^{0.04}$ & $160._{-10.(20.)}^{20.(10.)}$ \\
$  2.27$ & $0.22_{-0.06}^{0.02}$ & $180._{-10.(20.)}^{10.(10.)}$ \\
$  2.29$ & $0.2_{-0.04}^{0.02}$ & $186._{-10.(20.)}^{6.(10.)}$ \\
$  2.31$ & $0.18_{-0.03}^{0.02}$ & $190._{-10.(20.)}^{5.(10.)}$ \\
$  2.33$ & $0.18_{-0.02}^{0.02}$ & $188._{-10.(20.)}^{4.(10.)}$ \\
$  2.375$ & $0.16_{-0.01}^{0.02}$ & $192._{-10.(20.)}^{4.(10.)}$ \\
$  2.42$ & $0.14_{-0.01}^{0.01}$ & $190._{-9.(20.)}^{5.(10.)}$ \\
$  2.465$ & $0.13_{-0.01}^{0.01}$ & $192._{-8.(20.)}^{6.(10.)}$ \\\bottomrule   
\end{tabular}

\bibliographystyle{apsrev4-1}
\bibliography{bibliography,new}
\end{document}